\documentclass[%
superscriptaddress,
 amsmath,amssymb,
 aps,
 prb,
reprint,
floatfix,
]{revtex4-2}
\usepackage{graphicx,xcolor}
\definecolor{darkblue}{RGB}{0,0,150}
\definecolor{nightblue}{RGB}{0,0,100}

\usepackage{mathrsfs,dsfont,mathtools}

\usepackage{dcolumn}
\usepackage{bm}
\usepackage[
colorlinks,
citecolor=darkblue,
linkcolor=darkblue,
urlcolor=nightblue]{hyperref}

\usepackage[english]{babel}
\usepackage[babel,kerning=true,spacing=true]{microtype}

\usepackage{feynmp-auto}

\definecolor{DarkRed}{RGB}{100,0,0}
\definecolor{DarkBlue}{RGB}{000,0,100}

\AtBeginDocument{\renewcommand{\natexlab}[1]{#1}}
\begin{document}

\title{Even Integer Quantum Hall Effect in Materials with Hidden Spin Texture}

\author{Daniel Kaplan}
\email{d.kaplan1@rutgers.edu}
\affiliation{Department of Condensed Matter Physics,
Weizmann Institute of Science,
Rehovot 7610001, Israel}
\affiliation{Department of Physics and Astronomy, Center for Materials Theory,
Rutgers University, Piscataway, NJ 08854, USA}
\author{Ady Stern}
\affiliation{Department of Condensed Matter Physics,
Weizmann Institute of Science,
Rehovot 7610001, Israel}
\author{Binghai Yan}
\affiliation{Department of Condensed Matter Physics,
Weizmann Institute of Science,
Rehovot 7610001, Israel}
\begin{abstract}
Because spin-orbit coupling (SOC) is invisible in the band structure when inversion symmetry exists, whether spins are trivially degenerate or strongly coupled to momentum due to SOC is presumed to make little difference in transport measurements, such as magnetoresistance and quantum oscillations.
In this work, however, we show that hidden Rashba SOC in a centrosymmetric two-dimensional material can lead to the quantum Hall effect with only even-integer plateaus, unlike a spinless electron gas. 
Here, two Rashba layers that are degenerate but with opposite SOC due to inversion symmetry, hybridize with each other and create two doubly-degenerate bands with hidden spin texture. Correspondingly, two branches of Landau levels interact, resulting in significant suppression of spin splitting due to the balancing of intralayer SOC and interlayer hybridization. 
Furthermore, we show that breaking inversion symmetry restores the ordinary quantum Hall fluid by introducing spin-split Fermi surfaces. Our theory can apply to centrosymmetric materials with strong SOC, as demonstrated in a recent experiment on the two-dimensional semiconductor Bi$_2$O$_2$Se.
\end{abstract}

\maketitle

\section{Introduction}
The study of electronic properties in a magnetic field is essential for characterizing and understanding electrons on a Fermi surface (fermiology) \cite{Alexandradinata2023_fermiology,Ziese2020,ashcroft2022solid}. Classically \cite{ashcroft2022solid}, electrons subjected to a magnetic field follow circular orbits that begin to drift in the presence of an electric field. An important consequence in a two-dimensional electron gas (2DEG) is the Hall effect which presents as the emergence of a transverse voltage (or current in an infinite system) in response to a longitudinal electric field. Quantum mechanically, the 2D limit leads to the quantization of electronic orbits leading to the quantum Hall effect \cite{Klitz1980,Laughlin1981,Thouless1982,Klitz2020}. This remarkable -- and the first identified -- topological phase of matter consists of an insulting bulk and chiral edge states on the boundary of the 2D system. In the quantum Hall effect, the longitudinal conductance $\sigma_{xx}$ vanishes while the transverse conductivity $\sigma_{xy}$ appears as plateaus (versus magnetic field), precisely quantized at integer values of $e^2/h$ \cite{Klitz1980}. 

Theoretically, the problem is formulated as the motion of electrons in a parabolic band (with an effective mass)  \cite{Landau1930} that interact with the magnetic field through a gauge description and minimal coupling. The resultant quantization is a spectrum of states (known as Landau levels (LLs) \cite{Sakurai1967}) that are completely flat (momentum independent) and thus have no dispersion and are incompressible. The energies of these states are separated by integer quanta of the classical Larmor frequency $\omega_c = \frac{e B}{m^*}$ where $B$ is the magnetic field and $m^{*}$ is the effective electron mass.  Real materials and electronic dynamics add additional degrees of freedom to the electronic motion. One such degree of freedom is electronic spin and the coupling between spin and the Zeeman effect expressed through the $g_0$-factor. Whenever the Zeeman energy (defined through $g_0$) is weaker than the typical disorder strength, 

LLs are doubly degenerate with respect to spin (for example in graphene \cite{novoselov2006unconventional,novoselov2007room,li2007observation}). As a result, Hall plateaus occur at steps which are $f$-fold apart, where $f$ stands for the degeneracy \cite{lifshits1958theory}. For ultraclean graphene samples \cite{Zhang2006_graphene}, a magnetic field dependent splitting of Hall plateaus demonstrated a Zeeman-like lifting of LL degeneracies. 
The existence of spin may involve 
spin-orbit coupling (SOC), a common feature of semiconductors \cite{spintronics_review_dassarma}. Traditionally, the importance of SOC to magnetotransport was realized in GaAs-based systems \cite{Das1989,Junsaku1997,Ikai2002,manfra2014molecular}. There, the lack of inversion symmetry allowed the explanation of the $B \to 0$ splitting observed in Shubnikov-de Haas (SdH) experiments via the Rashba/Dresselhaus effects \cite{bercioux2015quantum}. To leading order in momentum, the Rashba effect is given by a linear coupling of spin and electronic momentum that preserves rotational symmetry and thus lifts the degeneracy of spin states at finite momentum even without the Zeeman effect. Such splitting relies on inversion symmetry breaking and is thus not expected in centrosymmetric crystals \cite{winkler2003spin}.

\begin{figure}
    \centering
    \includegraphics[width=\columnwidth]{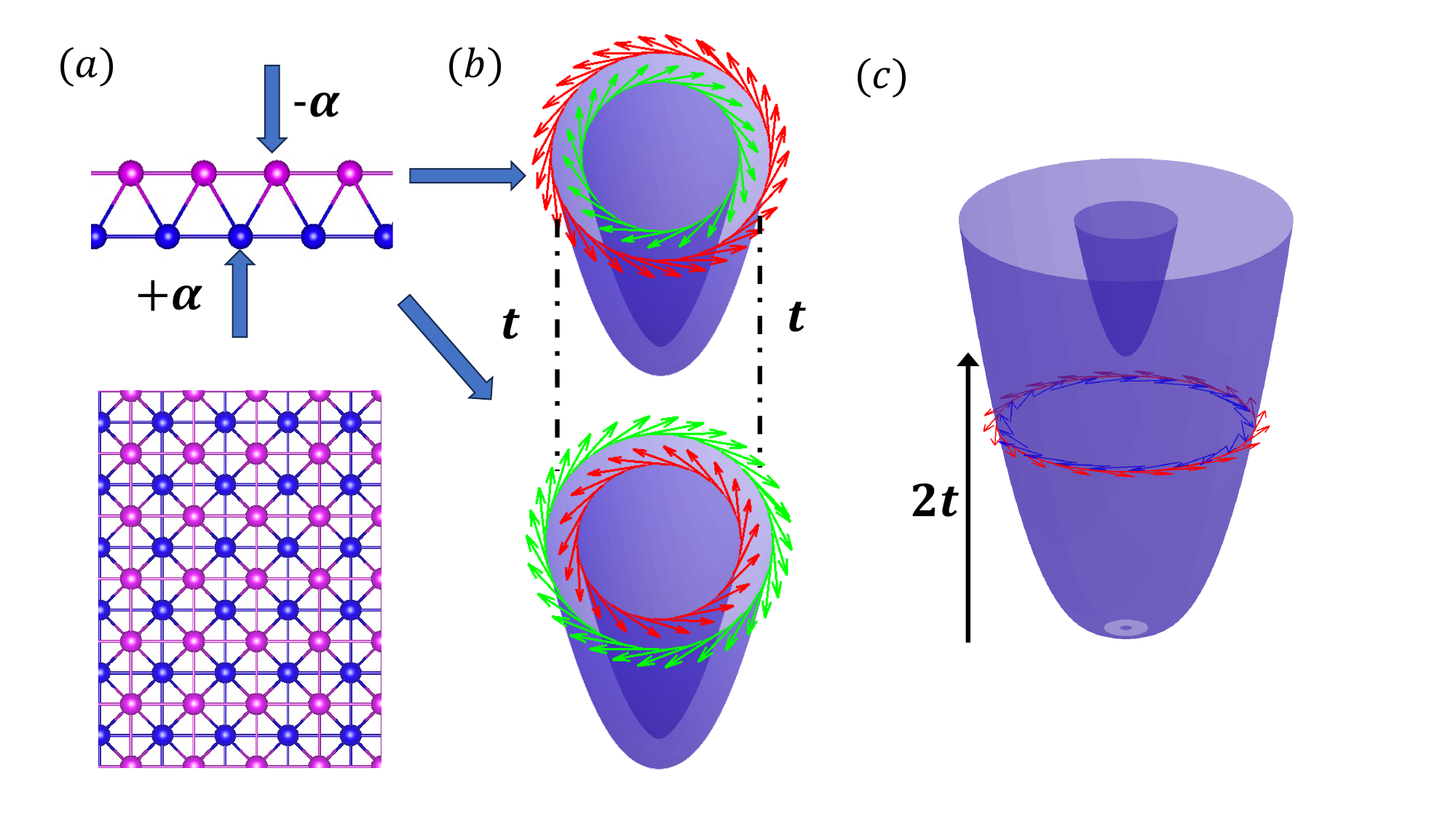}
    \caption{Illustration of the hidden SOC effect. (a) A checkerboard lattice (side view, above; top view, below) where each sublattice can support a finite dipole. Since the overall structure retains inversion symmetry, the total dipole moment is zero. Each sublattice has an effect SOC strength $\pm \alpha$. (b) Without coupling, each sublattice is assumed to have a quadratic dispersion, leading to spin-split Fermi surfaces (red and green, with spin projection superimposed). The coupling is taken to have strength $t$. (c) After the coupling is introduced, for energies $E < 2t$, a single, doubly degenerate Fermi surface emerges, with no net spin polarization thus leading to hidden spin texture. }
    \label{fig:fig1}
\end{figure}

Recently, there has been considerable interest in centrosymmetric crystals exhibiting ``hidden" SOC effects \cite{zhang2014hidden,yuan2019uncovering,Guan2023}. In a system with hidden SOC, the global crystal structure has inversion symmetry, but individual constituents or subsets of the structure may carry a finite dipole field (due to local site asymmetry), while the overall field is constrained to be zero by inversion symmetry. The dependence on the local site symmetry is naturally understood as arising from SOC being an atomic effect \cite{zhang2014hidden}, which is predominantly sensitive to the local environment. To make this concrete, consider a lattice with strong atomic SOC, and specifically a two dimensional checkerboard lattice (which is a 2D analogue of a 3D pyrochlore lattice) \cite{hu2023realization}. In Fig.~\ref{fig:fig1}(a) we show an example of two sub-units/layers, related to each other by inversion symmetry. For example, one can imagine a checkerboard lattice with sublattices coupled together. Each unit has site-asymmetry allowing it to support a finite dipole moment. The overall structure retains inversion symmetry, and therefore imposes the condition that the overall dipole charge in the system is zero. 

Individually, every subunit's Fermi surface has spin splitting due to inversion symmetry breaking with a momentum-locked spin texture, which is opposite in projection on the other subunit; this is shown in Fig.~\ref{fig:fig1}(b). Considered together, and allowing for hybridization, the full system recovers inversion symmetry, in which combined with time-reversal symmetry guarantees that the energy bands are (at least) doubly degenerate at every (Fig.~\ref{fig:fig1}(c). This is so, since inversion inverts the layer/sublattice degree-of-freedom (DOF) and momentum but leaves spin unchanged, while time-reversal only inverts momentum and spin. The combined operation is a symmetry of the structure, is local in momentum and creates a doubly-degenerate spectrum at every $\mathbf{p}$.

Due to the abundance of centrosymmetric structures that do not carry an overall dipole moment, it is natural to ask whether a signature in magnetotransport would reveal the hidden spin texture and the effect of SOC, beyond direct (and difficult) probing of the band structure using ARPES and optical instruments \cite{wu2017direct,Razzoli2017,clark2022hidden,hu2023realization}.

In this work, we answer this question by calculating and predicting unique mangetotransport in inversion symmetric systems with hidden Rashba SOC. We show that unlike the SOC in a normal parabolic band (which always results in the lifting of spin degeneracy), the hidden spin texture suppresses the Zeeman splitting of Landau levels. We demonstrate that the suppression is due to competition between the hybridization strength $t$, the atomic SOC strength $\alpha$ and the Zeeman energy proportional to $g_0$ (the bare g-factor). The presence of two subunits splits the LLs and the SOC into two species. SOC can only couple LLs of different subunits leading to the energy splitting being suppressed by $t^{-1}$ which becomes exactly zero in the limit of strong hybridization. 

We present a remarkable consequence of this competition: for a range of SOC strength, the effective splitting of Landau levels can be tuned to zero leading to a degeneracy between electronic orbits that persists to arbirarily high magnetic fields.
As a consequence, magnetotransport experiments will exhibit a Hall conductivity with plateaus which occur at \textit{even} steps, with $\sigma_{xy}$ now a multiple of $2e^2/h$. This is entirely unlike the usual effect of SOC in semiconducting systems which leads to $\sigma_{xy}$ to be an integer multiple of $e^2 /h$ showing both even and odd plateaus. 

As the effect relies on the inversion-symmetry-protected hidden spin texture, we show how breaking inversion symmetry (e.g., by an out-of-plane electric field or a substrate) will lead to the restoration of odd-integer plateaus to the Hall conductivity and a splitting of the fundamental frequency in quantum oscillations indicating the existence of two Fermi surfaces. We conclude by presenting a material example -- the layered high-mobility semiconductor Bi\textsubscript{2}O\textsubscript{2}Se \cite{wu2017high} which realizes this unique spin polarization with favorable SOC strength and interlayer coupling \cite{Wang2024}. 

Our findings in hidden SOC systems will be important for future spintronic \cite{spintronics_review_dassarma} and orbit-tronic device \cite{go2021orbitronics} applications and also for recent experiments \cite{sheng2021rashba}, offering new platforms for the observation of unusual transport characteristics and quantum Hall related phenomena.
This paper is organized as follows. In Sec.~\ref{sec:singleband} we review the effects of SOC on Landau quantization in a single band. In Sec.~\ref{sec:spintext} we introduce the theoretical formalism for Landau quantization with hidden spin polarization. Here, we discuss at length the different regimes of the coupling between the two sublattices/subunits and the physical consequence on experimental findings. In Sec.~\ref{sec:magneto} we outline the method for computing transport phenomena: quantum oscillations inferred from chemical potential fluctuation as a function of magnetic field and the quantized Hall conductivity. In Sec.~\ref{sec:matexam} we suggest materials candidates, and perform first-principes calculation on 4 unit cell Bi\textsubscript{2}O\textsubscript{2}Se, offering the possibility of measuring the unique Hall transport predicted in this work.

\section{Single band Landau quantization with SOC}
\label{sec:singleband}

\begin{figure}
    \centering
    \includegraphics[width=\columnwidth]{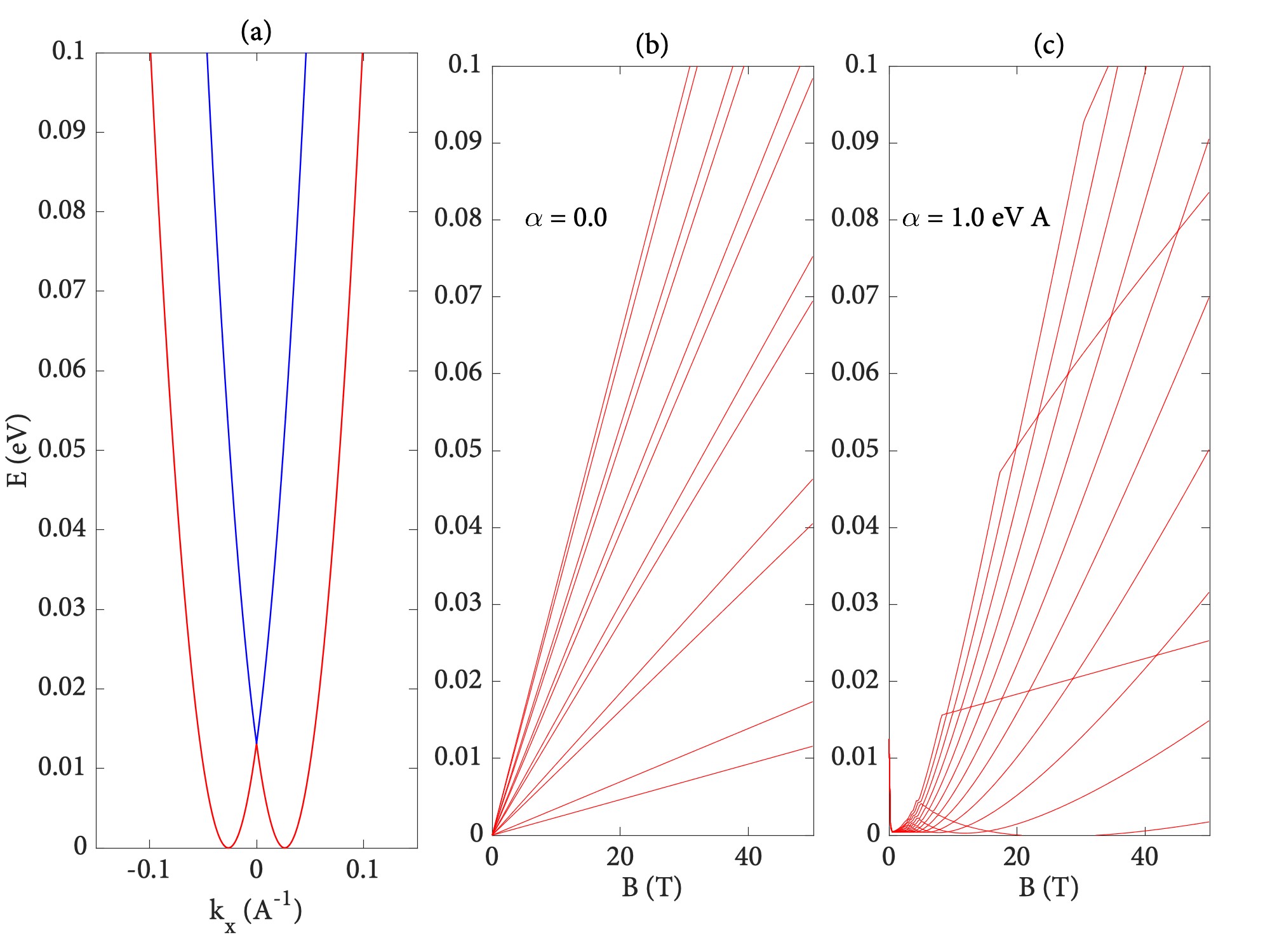}
    \caption{Dispersion and Landau levels for a single parabolic band with SOC. (a) Dispersion along $k_x$ showing the formation of two Fermi surfaces for any energy $E$. Red and blue denote the energy bands $\varepsilon_{p}^{\pm}$, respectively. (b). Landau fan with zero SOC, showing Landau levels split by the Zeeman energy $E_z = g_0 \mu_B B$. Splitting is Landau level (LL) independent. (c) LLs with $\alpha = 1.0 \mathrm{eV \AA}$. Significant Landau level crossings can be obesrved, and the level splitting differs substantially from $E_z$.}
    \label{fig:main_fig2}
\end{figure}

In this section, we summarize the main findings concerning single band dynamics in the presence of a magnetic field. The full derivation and analysis of the single band model is presented in App.~\ref{app:singleband}.
We start with a single parabolic band dispersion with Rashba-type SOC for a particle with effective mass $m^{*}$, and arrive at the Hamiltonian,
\begin{align}
    H = \frac{p^2}{2m^{*}} + \frac{\alpha(\sigma_y p_x - \sigma_x p_y)}{\hbar},
    \label{eq:single_band_H}
\end{align}
Here $p^2 = p_x^2 + p_y^2$. This dispersion is plotted in Fig.~\ref{fig:main_fig2}(a). The energies are given by,
\begin{align}
    \varepsilon_{p, \pm} = \frac{p^2}{2m^{*}} \pm \frac{\alpha}{\hbar} p.
\end{align}
The density of states for this system is given by,
\begin{align}
    \mathcal{D}(E, \varepsilon_{\pm}) = \frac{2 \pi m^{*}} {\hbar^2} \left | \frac{\alpha}{\hbar \sqrt{\alpha^2 \hbar^{-2} + 2E/m^{*}}} \pm 1\right|,
\end{align}
for all $E \geq -m^{*}\alpha^2/\hbar^2$.

The immediate result observed here is that a new energy minimum with a \textit{divergent} density of states appears at $E = - m^{*}\alpha^2 \hbar^{-2}/2$. The enhanced density of states is the source of the Landau levels appearing near zero energy in Fig.~\ref{fig:main_fig2}(c). These states vanish in the limit of $\alpha \to 0$ where the standard $D(E) \sim \textrm{const}$, 2D density of states is recovered giving the usual Landau spectrum found in Fig.~\ref{fig:main_fig2}(b).
We quantize the problem with a magnetic field in the Landau gauge and add Zeeman spin coupling $H_z = \frac{g \mu_B B}{2} \sigma_z$. The Hamiltonian reads,
\begin{align} 
\notag
    H &= \frac{1}{2m^{*}} \left(p_x^2+(p_y+e B x)^2\right) + \\ & \frac{\alpha}{\hbar} \left(\sigma_y p_x -\sigma_x (p_y+e B x) \right) + \frac{g \mu_B B}{2} \sigma_z.
    \label{eq:main_singleband_ham}
\end{align}
The eigenvalues of this Hamiltonian are given by the spectrum,
\begin{align}
    \notag \varepsilon_{n, \sigma= \pm 1} &= \left(n+\frac{1}{2}(1+\sigma)\right)\hbar \omega_c -\\ &  \sigma \sqrt{\frac{2(n+(1+\sigma)/2)\alpha^2}{l_B^2} + \left(\frac{E_z}{2}\right)^2\left(1-\frac{2 m_0}{g m^{*}}\right)^2}.
    \label{eq:main_spectrum_singleband}
\end{align}
For compactness, we denoted $E_z = g \mu_B B$, which is the Zeeman energy and $\omega_c = \frac{eB}{m^{*}}$. The magnetic length is $l_B^2 = \hbar (eB)^{-1}$. The detailed analysis of the solution is discussed in the appendix, App.~\ref{app:singleband}.

Here, we remind the reader the conclusions concerning the level splitting. The effective $g$ factor, which defines the linear in magnetic field dependent splitting of LLs, is given by, 
\begin{align}
    g_\textrm{eff} = \frac{\Delta \varepsilon}{\mu_B B} = \frac{\varepsilon_{n,\sigma} - \varepsilon_{n,\sigma'}}{\mu_B B}.
    \label{eq:main_gfactor}
\end{align}
In the case without SOC ($\alpha = 0$), the splitting for every Landau level $n$ grows linearly with field $B$, as expected from Zeeman splitting. When SOC exists, the levels mix, and the identification of the relevant energy states with the quantum numbers of the zero SOC case is no longer possible. The main result, however, is that the splitting of LLs never vanishes, and is in fact significantly enhanced, e.g., in the limit of $B \to 0$.

The energy splitting is evaluated for arbitrary SOC strength, from the exact solution. The relevant quantity is given by,
\begin{align}
\notag
\Delta \varepsilon &= \hbar \omega_c - \sqrt{2 (n+1)\alpha^2 l_B^{-2}+E_z^2 (1-q)^2/4} - \\ & \sqrt{2n\alpha^2 l_B^{-2}+E_z^2 (1-q)^2/4}.
\label{eq:main_full_split}
\end{align}

Analytically,the behavior in the vicinity of $\alpha \to 0$ is $\Delta \varepsilon \propto \alpha^2$ which can be understood from treating SOC perturbatively, in the basis of spin-split states with a finite $g_0$. Calculated directly to order $\mathcal{O}(\alpha^2)$ in the limit $\alpha l_B^{-1} \ll E_z$, we have,
\begin{align}
\Delta \varepsilon \approx E_z - \frac{\textrm{sgn}(q-1)2(2n+1)\alpha^2}{E_z l_B^2 |1-q|} + \mathcal{O}(\alpha^3).
\label{eq:main_perturb_level_splitting}
\end{align} 
In the opposite limit, i.e., $\alpha l_B^{-1} \gg E_z$ to order $E_z^2$,
\begin{align}
    \Delta \varepsilon  \approx \hbar \omega_c - C_n \alpha l_B^{-1} - \frac{C_n E_z^2 l_B}{16 \alpha\sqrt{n}\sqrt{n+1}},
\end{align}
with $C_n = \sqrt{2}\left(\sqrt{n}+\sqrt{n+1}\right)$.
The fundamental result is that, barring accidental degeneracies, for arbitrary values of $\alpha$ the LLs are energetically well-separated. This is true even if the bare g-factor $g_0 = 0$. In Sec.~\ref{sec:magneto} we explain the ramifications of this fact on magnetotransport and how this level splitting leads to the observation of ordinary (one quantum of $e^2/h$) Hall plateaus and single discernible peaks in quantum oscillations.

To investigate how the band structure affects spin-splitting, in the next section we consider a model which enforces degenerancies at the band structure level, leading to a renormalized Landau spectrum and effective spin splitting.

\section{Landau quantization with hidden spin texture}
\label{sec:spintext}

\subsection{Band structure}
The double Rashba layer model includes two coupled Rashba monolayers with opposite SOC. The simplest model realizing structural inversion and enforcing double-degeneracy at every $k$-point in the spectrum is a four-band model with spin and layer DOFs. With SOC, inversion symmetry is preserved when the sign of the SOC is inverted between layers, realizing a hidden-Rashba effect.

\begin{figure}[ht]
    \centering
    \includegraphics[width=\columnwidth]{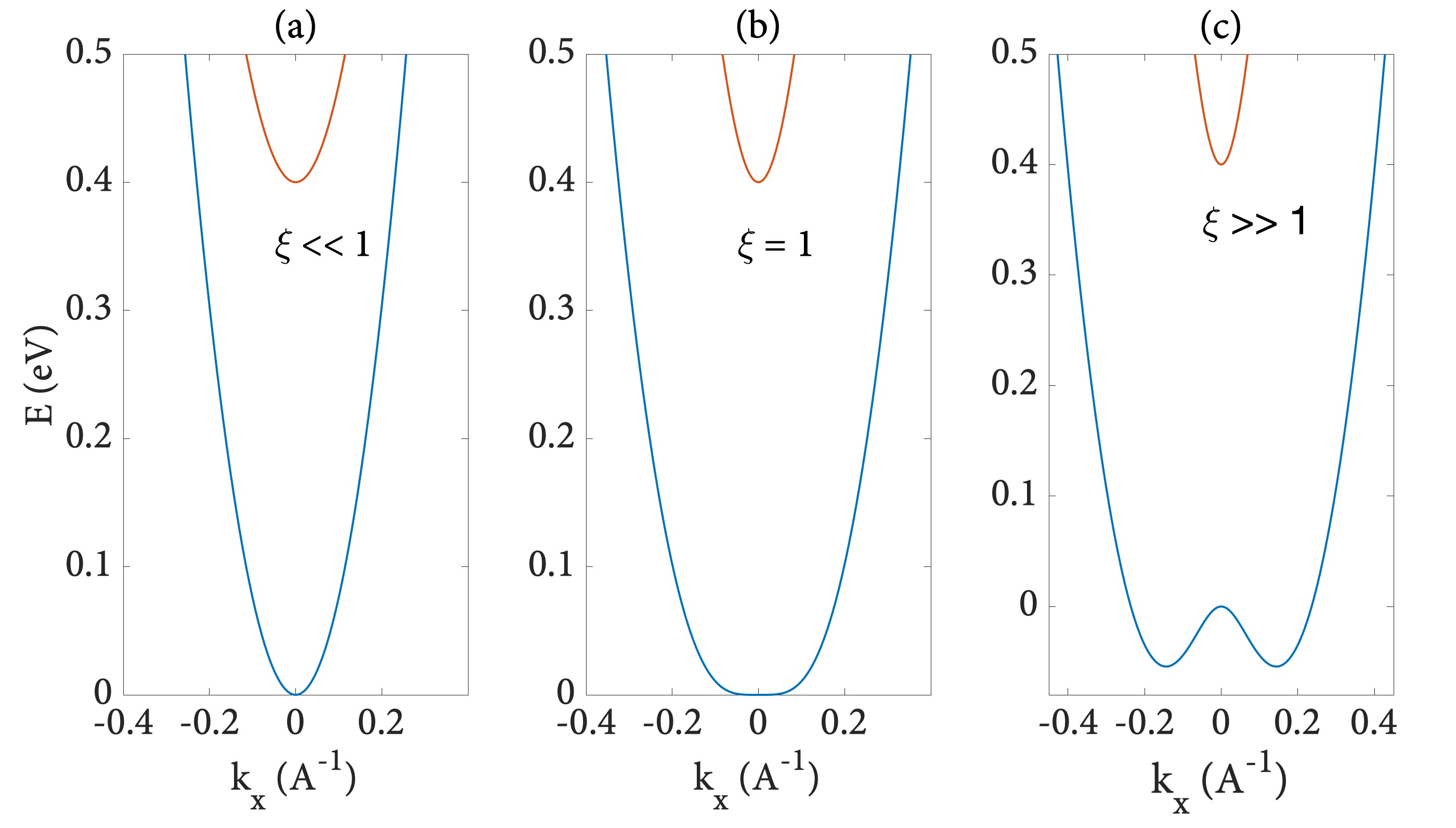}
    \caption{Band structure for the system with hidden spin texture. All band structures are plotted at $k_y = 0$. (a) Bands for vanishing SOC strength, showing two bands separated by $2t$ (hybridization gap) and unrenormalized effective mass $m^{*}$ (Eq.~\eqref{eq:effec_mass}). (b) Band structure for the system with a critical value of SOC corresponding to $\xi = 1$. The parabolic part of the dispersion vanishes for the bottom band and $\varepsilon \sim p^4$ is the leading order contribution, as in Eq.~\eqref{eq:quartic_disp}. (c) For $\xi \gg 1$, two minima appear in the dispersion. This case can be distinguished from the former, by the appearance of two Fermi surfaces at low densities.}
    \label{fig:fig4}
\end{figure}

We consider a model for electrons with effective mass $m^{*}$ in two identical layers/sublattices with opposite SOC. Denoting the layer/sublattice degree of freedom by the Pauli matrices $\tau$, and spin being $\sigma$, we write,
\begin{align}
	H = \frac{p^2}{2m^{*}} - \frac{\alpha}{\hbar} \tau_z (p_x \sigma_y - p_y\sigma_x) -t\tau_x.
\end{align}
Here, the layers are hybridized by the parameter $t$. At this level, a new dimensionless scale emerges to quantify the SOC strength relative to interlayer hybridization,
\begin{align}
\xi = \frac{m^{*} \alpha^2}{\hbar^2 t},
\end{align} 
which captures the effect on the band structure and the behavior of the density of states near $\Gamma$ through a dimensionless quantity. \begin{align}
	\varepsilon_{\sigma,\tau=\pm} = \frac{p^2}{2m^{*}} \pm \sqrt{\frac{\alpha^2 p^2}{\hbar^2} + t^2} = \frac{p^2}{2m^{*}} \pm t \sqrt{1+ \frac{\xi p^2}{t m^{*}}}.
\end{align}
The spectrum for every value of the layer index $\tau$ is doubly degenerate, as expected from a system with combined inversion $\mathcal{P}$ and time-reversal $\mathcal{T}$ symmetries. Inversion is represented in the model by $\mathcal{P} = \tau_x (p \to -p)$, and time-reversal $\mathcal{T} = i\sigma_y \mathcal{C}(p\to -p)$ ($\mathcal{C}$ is complex conjugation). The combined symmetry commutes with $H$ as well, allowing for the double degeneracy for every momentum value, as it is local in $p$. 
We focus on the lower band, indexed by $\tau = -1$. The inverse effective mass for this band reads,
\begin{align}
    \frac{1}{m} = \frac{1}{m^{*}} - \frac{\xi t}{(m^{*}t +p^2 \xi)\sqrt{1 + p^2 \xi / m^{*}t}}.
\end{align}
Moreover, at $\Gamma$,
\begin{align}
    \frac{1}{m_\Gamma} = \frac{1}{m^{*}}(1- \xi),
    \label{eq:effec_mass}
\end{align}
vanishes identically for $\xi = 1$. In the vicinity of the $\Gamma$ point then, the dispersion becomes flat,
\begin{align}  
\varepsilon_{\sigma,\tau=-1}(\xi = 1) = \frac{p^4}{8m^2t} + \mathcal{O}(p^6).
\label{eq:quartic_disp}
\end{align}
The density of states then becomes,
\begin{align}
    D(E) = \frac{2\pi \sqrt{2m^2 t}}{\hbar^2\sqrt{E}},
\end{align}
which diverges at $E \to 0$ with consequences for the Landau spectrum as presented in the next section. In Fig.~\ref{fig:fig4} we plot the dispersion of the inversion symmetric cell as a function of $\xi$. We keep $m^{*}, t$ fixed throughout. Small values of $\xi$ thus correspond to weak atomic SOC. Large values, $\xi \gg 1$ are those for which additional minima appear in the energy spectrum as shown in Fig.~\ref{fig:fig4}(c). All bands are doubly degenerate due to $\mathcal{P}\mathcal{T}$ symmetry.
\subsection{Landau levels with inversion symmetry}
We now apply the Landau gauge to the operators of the two layer problem. Using the same replacement $\mathbf{p} \to \mathbf{p} + e\mathbf{A}$, the Hamiltonian in the layer and LL basis $\tau_z = \pm 1$ takes the form,
\begin{align}
    \notag &H_{nm}^{ll'} = \left[\hbar \omega_c \left(n + \frac{1}{2}\right)\mathbf{1} + \frac{E_z}{2} \sigma_z\right]\delta_{nm} \delta_{ll'} - \\ \notag & \textrm{sgn}(l) \delta_{ll'}\frac{\sqrt{2} \alpha}{l_B} \left(\sqrt{m} \delta_{n,m-1} \sigma^{+}+\sqrt{m+1}\delta_{n,m+1} \sigma^{-}\right) - \\ & t\delta_{l,l'\pm 1} \delta_{nm}.
\label{eq:fullham_layerbasis}
 \end{align}
In this form, the Hamiltonian for each layer index $l$ is a $2 \times 2$ matrix in the spin-basis, similar to the one band model. In the limit $\alpha \to 0$, this analogy becomes exact, and the two $l$-indexed layers decompose into two quantum Hall fluids with different SOC couplings. Since the coupling enters the solution quadratically (Eq.~\eqref{eq:main_spectrum_singleband}), the two quantum Hall states retain exactly identical spectra and hence splitings, within their respective subspace. When the coupling $t$ is switched on, the SOC no longer commutes with the layer basis. In the limit of $t \gg E_z, \hbar \omega_c, \alpha l_B^{-1}$, the layers are fully hybridized and the SOC enters perturbatively, with the leading correction appearing at 2nd order. 

The exact eigenvalues of Eq.~\eqref{eq:fullham_layerbasis} can be obtained by observing that the coupling term makes $H$ tridiagonal in the layer basis. Using techniques similar to those used for the spin basis in the one-layer case, we find the spectrum to be given by,
\begin{align}
    \varepsilon_{n,\sigma,\tau} = n_\sigma \hbar \omega_c +\tau \sqrt{2 n_\sigma \alpha^2 l_B^{-2} + ( t + \sigma(E_z -\hbar\omega_c)/2)^2}.
    \label{eq:twobandsol}
\end{align}
Here, we defined $n_\sigma = n+(1+\sigma)/2$. In this expression the original DOFs of layer $\tau = \pm 1$ and spin $\sigma = \pm 1$ are fully hybridized. As expected, in the $t \to 0$ limit it reduces to Eq.~\eqref{eq:main_spectrum_singleband} (with the relabelling of $\tau \leftrightarrow \sigma$), the additional degeneracy being due to the fact that $[H, \tau_z] = 0$ rendering $\tau$ a good quantum number. Note that unlike the case without magnetic field, the Landau levels are not doubly degenerate, as the Zeeman term does not commute with $\mathcal{P}\mathcal{T}$. However, as we show below, tuning the SOC and inter-layer coupling suppresses Landau
level splitting.
\subsection{Small $t$}
In the absence of any coupling between the layers, we find that the solution of two states becomes degenerate in their lower energy sub-index. In this case, the smallest energy scale for LL splitting is dictated by $t$. 
The leading order correction is found by expanding Eq.~\eqref{eq:twobandsol} in $t \ll E_z, \hbar\omega_c, \alpha l_B^{-1}$, and relabelling $\tau$ with $\sigma$. Here, the energy levels take the form (to leading order in $t$),
\begin{align}
    \varepsilon_{n, \sigma, \tau} \approx n_\sigma \hbar \omega_c + \sigma \tilde{E} + t \frac{\sigma \tau (E_z - \hbar \omega_c)}{\tilde{E}},
\end{align}
where we defined $\tilde{E} = \sqrt{2n_\sigma \alpha^2 l_B^{-2} + ((E_z -\hbar\omega_c)/2)^2}$. The solutions are degenerate in the $\tau$ index in the limit of $t \to 0$, as explained above. Qualitatively the result can be understood as coming from two identical LL species with $t$-scale splitting by the intermixing of the $\sigma, \tau$ DOFs. An inspection of the dispersion reveals the reason: while the additional degree of freedom ($\tau$) doubles the number of bands, the weakness of the hybridization still produces the same double-well energy landscape, due to the SOC. The result is a nearly vanishing splitting for the same spin specie, i.e. $\Delta \varepsilon \sim 2t$ in the leading order, but two nearly degenerate Fermi surfaces, for a fixed density. Hence, a vanishing $t$ case would correspond to nearly degenerate Landau levels, but with the measurement of two frequencies in the observed quantum oscillations due to the lifting of the degeneracy in the $\sigma$ channel. 

\subsection{Large $t$}

The large $t$ limit occurs in a material where the hybridization energy of orbitals (e.g. $p$ orbitals) is the dominant energy scale. This is a common feature of layered semiconductors. This implies the set of constraints $t \gg \hbar \omega_c, \alpha l_B^{-1}, E_z$. In the material example we provide (Sec.~\ref{sec:matexam}), this scale assumes the value $t \sim 0.2\textrm{eV}$. For an effective electron mass $m^{*} = 0.2m_0$, the cyclotron energy is then $\hbar \omega_c \sim 30 \mathrm{meV}$ (at $B = 50 \mathrm{T}$) with the Zeeman splitting $E_z \sim 5 \mathrm{meV}$. The SOC splitting at $B=50\mathrm{T}, \alpha = 1.5 \mathrm{eV \AA}$ is $\alpha l_B^{-1} \sim 40 \mathrm{meV}$. Thus, for many semiconducting systems, even with sizable on-site SOC strength, the dominant energy scale remains the hybridization energy.
The Landau spectrum is then constructed on top of the initially $\sim 2t$ split bands at the $\Gamma$ point. By expanding Eq.~\eqref{eq:twobandsol} in the limit $t \to \infty$,the energy levels have the form,
\begin{align}
    \notag \varepsilon_{n,\sigma,\tau} &= \tau t + \left(n_\sigma \hbar \omega_c + \frac{\sigma\tau}{2}(\hbar\omega_c-E_z)\right) + \\  & \frac{n_\sigma \alpha^2 \tau}{l_B^2 t} - \frac{\sigma \tau n_\sigma \alpha^2 (E_z-\hbar \omega_c)}{2l_B^2 t^2} + \mathcal{O}(t^{-3}).
\end{align}
As expected, the leading order term, $\pm t$, is simply the band splitting driven by the hybridization at $\Gamma$. The next to leading order is the Landau spectrum emergent above the scale $t$. Note that in this limit, the SOC is suppressed completely at the leading order (see Dicussion). The resultant Landau spectrum consists of Zeeman split ordinary Landau levels. The SOC appears now only at order $\tau^{-1}$ realizing a renormalized SOC energy scale $\tilde{E}_{\textrm{SOC}} = \alpha^2 l_B^{-2}t^{-1}$. 

The emergence of this scale can be understood in the following manner. For large $t$, the expectation value of $\tau_z$  must necessarily vanish. For infinite $t$ the electron spends half of its time in each layer and the sign of the SOC is opposite in the two layers, so the effect it feels of the SOC averages out. For large $t$, it is therefore proportional to $1/t$. The leading order contribution (by second-order perturbation theory) must be \textit{off-diagonal}, i.e. $|\langle \tau=-1 | H_{\textrm{SOC}} | \tau=+1\rangle|^2$, connecting the two different LL species that are separated by $\Delta \varepsilon = 2t$ in energy. This indicates that the SOC mechanism in hidden-Rashba systems intermixes LL of bands that are separated by orbital hybridization. As this contribution is now \textit{quadratic} in the SOC energy scale, i.e., $(\alpha l_B^{-1})^2$ it is thus \textit{linear} in $B$ and enters into direct competition with the Zeeman term.
Note that the sign of this term is now entirely independent of the spin orientation or the sign of $\alpha$, as it is proportional to $\alpha^2$. At order $t^{-2}$, the Zeeman split and zero-point shift of the Landau levels $E_z-\hbar \omega_c$ are also renormalized by SOC, with the scale now dictated by $\alpha^2 l_B^{-2} t^{-2}$ which is a dimensionless parameter relating SOC strength at the induced cyclotron orbit to the hybridization energy.

The renormalized energy splitting can now be calculated using the same expansion. Specifically, we consider the splitting connected with varying Landau level index and spin. For branch $\tau$, up to order $t^{-1}$ we find,
\begin{align}
\Delta \varepsilon =  \varepsilon_{n, +, \tau} - \varepsilon_{n, -, \tau} =\frac{\alpha^2 \tau}{l_B^2 t}- E_z \tau + \hbar\omega_c (1+ \tau).
        \label{eq:energydifft}
\end{align}
Eq.~\eqref{eq:energydifft} is one of the central results of this work. It demonstrates that to leading order in $t^{-1}$ in the limit of large $t$, the splitting of Landau levels coupled with SOC sensitively depends on the layer hybridization $t$ and magnitude of SOC. Moreover, since the SOC is layer dependent in this model, the contribution of the cyclotron frequency to the splitting becomes layer dependent. To make this concrete, consider the lower branch where $\tau = -1$. In this case, $\Delta \varepsilon = -\frac{\alpha^2}{l_B^2 t} + E_z$. The cyclotron contribution vanishes identically (but is enhanced in the opposite branch, $\tau = +1$). However, when SOC is introduced, it enters with opposite sign to the leading order, Zeeman contribution. This can be understood as a direct consequence of the strong layer-SOC mixing. The product of $\tau_z \sigma_{x,y}$ enforces corrections to enter only at 2nd order, where they lower the ground state energy. Given that the leading order contribution enters linearly in $B$, we define a critical value $\alpha_{\textrm{cr}}$ for which the spin-splitting of the LLL (lowest Landau level) in the lower branch vanishes at order $t^{-1}$ (higher orders enter with higher powers of $B$ and are beyond Zeeman spltting).

By expanding to order $t^{-1}$ we find $\alpha_{\textrm{cr}}$ to be,
\begin{align}
   \frac{\alpha^2}{l_B^2 t} -E_z = 0~ \Rightarrow~  \frac{2 m_0 \alpha_{\mathrm{cr}}^2}{\hbar^2 t} = g_0.
   \label{eq:main1}
\end{align}
This is a central result of this work. This condition depends only on band structure properties of the electrons and is independent of the applied field. It is also independent of the Landau level index, since the splitting we consider is only of the originally Zeeman split Landau levels. 
The above condition can be related to the dimensionless parameter of the band structure as $\frac{2m_0}{m^{*}} \xi = g_0$. In the case of a dispersion of free electrons, such that $m^{*} = m_0$, one gets $2 \xi = g_0$. Further substituting $g_0 = 2$, sets $\xi = 1$. This means that Landau levels of free electrons become degenerate, at precisely the ratio of SOC to interlayer tunneling which flattens the lower band (as shown in Fig.~\ref{fig:fig4}(b)) and sends its mass to infinity.  
Corrections appear at order $t^{-2}$, which are sensitive to the Landau level index. However, in the limit of $t \to \infty$ but $\xi = \textrm{const}$ all higher order corrections vanish and the degeneracy condition for the Landau levels becomes exact. 
\begin{figure}[ht]
    \centering
    \includegraphics[width=\columnwidth]{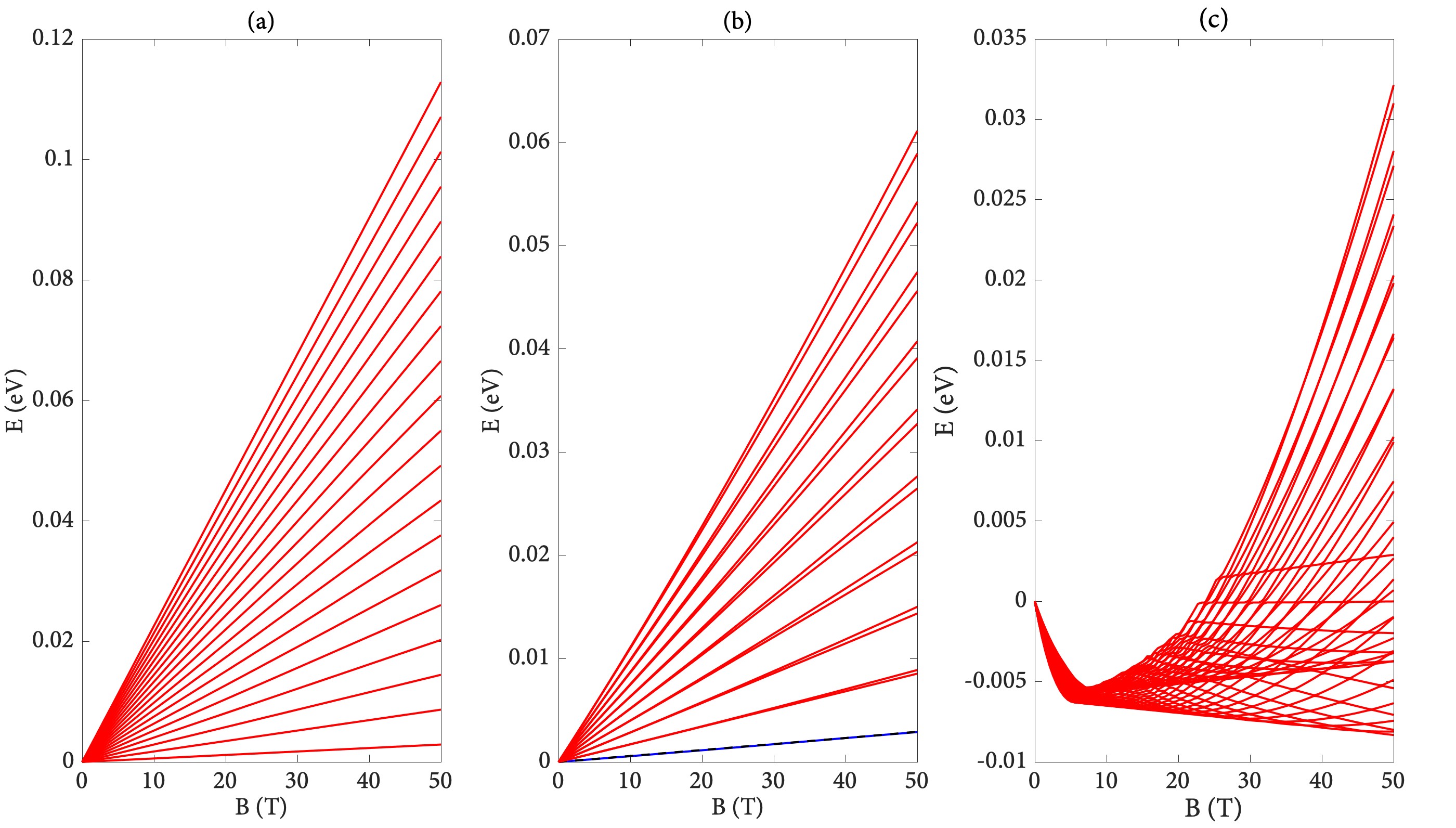}
    \caption{Landau levels in the presence of hidden spin texture. (a) Landau levels with vanishing SOC or $\xi \to 0$. Note that the Landau levels are seperated by the bare Zeeman splitting $E_z$ (and are therefore non-degenerate). The additional Landau fan that occurs at $E > 2t$ is not shown, but exhibits identical behavior. (b) Landau fan for the critical value $\xi = 1$. Note that the lowest Landau level has become effectively degenerate. We present the two spin species $\sigma = \pm 1$ as blue and black (dashed) lines, respectively. For $\xi = 1$, they overlap. Due to the higher-order corrections at order $t^{-1}$ which are Landau level dependent, the degenerancy is not exact for $n > 0$. (c) Landau fan for $\xi \gg 1$ showing significant crossings between bands of  different indices due to the added density of states at the new minima, as shown in Fig.~\ref{fig:fig4}(c). }
    \label{fig:fig5}
\end{figure}
In Fig.~\ref{fig:fig5} we show the consequences of the hidden spin texture on the Landau spectrum. Firstly, for weak SOC, the splitting is dictated by the Zeeman splitting, as expected. However, near the critical value of $\xi$ -- which depends on $\alpha$, and is extracted from Eq.~\eqref{eq:main1} this splitting is considerably suppressed. In the case of the LLL it can be tuned to nearly zero, as shown in Fig.~\ref{fig:fig5}(b). When $\alpha$ is very large, that is $\xi \gg 1$, the behavior observed in the single-band case is recovered and substantial Landau level crossing is seen. This is presented in Fig.~\ref{fig:fig5}(c). The degree of suppression can be quantified by extracting $g_\mathrm{eff}$ as we defined in Eq.~\eqref{eq:main_gfactor}. In Fig.~\ref{fig:fig6}, we plot $g_\mathrm{eff}$ for the inversion symmetric system, for a fixed magnetic field $B = 50 \mathrm{T}$ as a function of the site/sublattice SOC $\alpha$. Note the suppression of the LLL splitting at $\alpha \approx \alpha_{cr}$, inferred from Eq.~\eqref{eq:main1}. In Fig.~\ref{fig:fig6}(b) we also show that for increased $t$, the Landau-level dependent corrections which appear at order $t^{-2}$ in Eq.~\eqref{eq:main1} are significantly suppressed, as expected.
\begin{figure}[ht]
    \centering
    \includegraphics[width=\columnwidth]{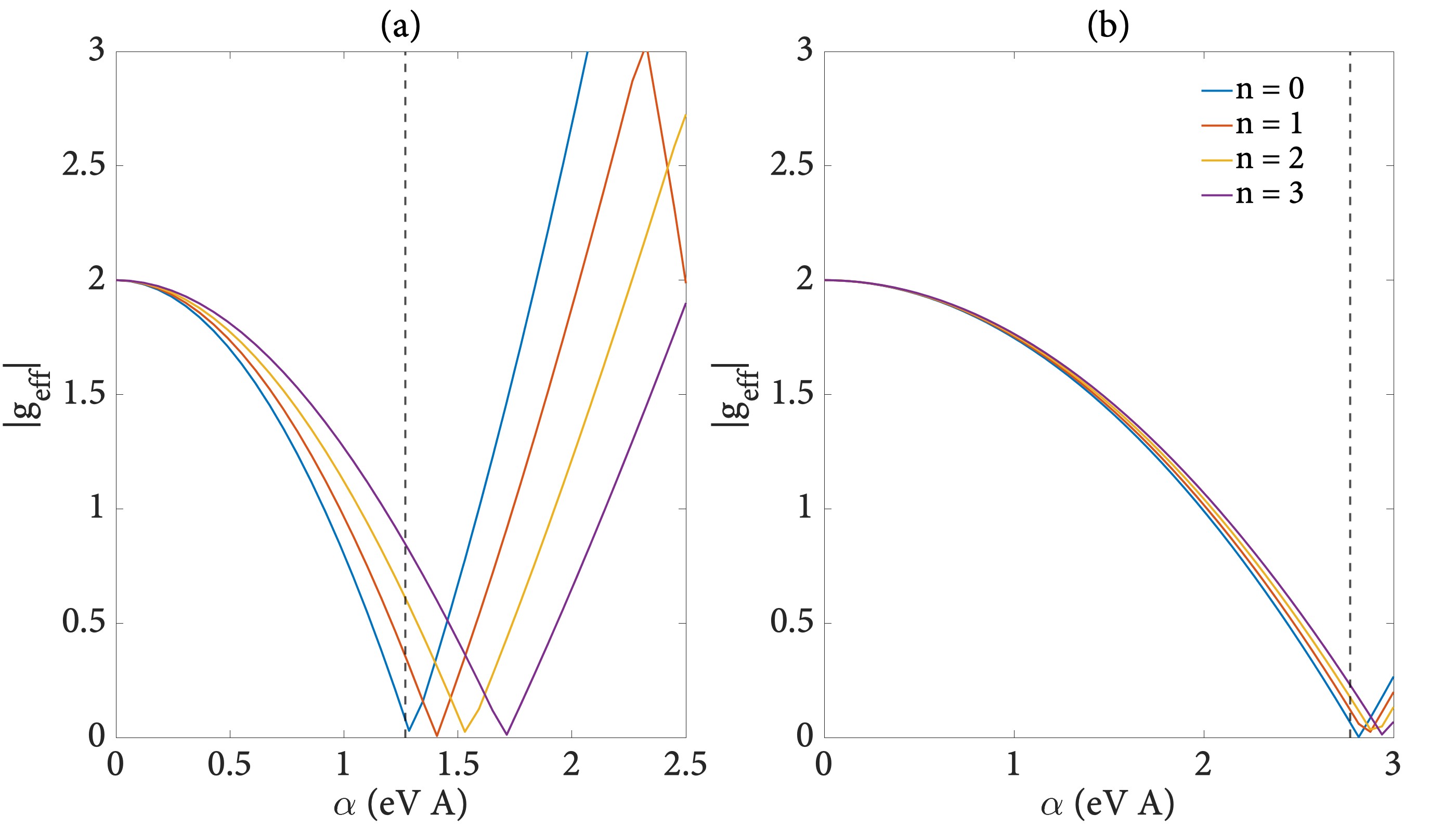}
    \caption{Landau level splitting as a function of SOC strength $\alpha$. For small $\alpha$, $g_\mathrm{eff} \to 2$, as expected. For values approaching the condition formulated in Eq.~\eqref{eq:main1}, the g-factor is strongly suppressed. In (a), $t = 0.2 \mathrm{eV}$, while in (b) $t = 1.0 \mathrm{eV}$. For $t \to \infty$, all higher-order corrections to the condition in Eq.~\eqref{eq:main1} vanish. Dashed line marks $\alpha_{\mathrm{cr}}$ determined from Eq.~\eqref{eq:main1}. For this figure, $B = 50 \mathrm{T}$.}
    \label{fig:fig6}
\end{figure}
The effect of the suppression of the $g$ factor on magnetotransport is reviewed in the next section.
\section{Magnetotransport}
\label{sec:magneto}
To understand the role of the suppression of $g$ factors and degeneracy of Landau levels we solve for the electronic conductivity $\sigma^{xy}$ and the density of states $\rho(E)$ for the Landau spectrum. $\sigma^{xy}$ holds the signatures of the quantum Hall effect, while $\rho(E)$ is directly related to oscillations in $\sigma_{xx}$ used to determine the Fermi surface via the Shubnikov- de Haas effect (SdH) \cite{lifshits1958theory}.
\subsection{Density of states and Hall conductivity}
Since the magnetic Hamiltonian always commutes with one of the momentum operators (in the Landau gauge), as shown in Eq.~\eqref{eq:singleband_ham}, the Landau problem is degenerate with the following number of states,
\begin{align}
N = \frac{S}{2 \pi l_B^2},
\end{align}
where $S$ is the area of the sample. We consequently define the density of states per unit area,
\begin{align}
    \rho(E) = \frac{1}{2 \pi l_B^2}\sum_{n,l}\delta(E-\varepsilon_n^{l}).
    \label{eq:density_states}
\end{align}
Here $l$ stands for the additional degrees of freedom for the Landau levels, i.e., $\sigma,\tau$, in the case of the bilayer system. As real systems have finite disorder, it is common to introduce broadening to the density of states. We broaden the delta function of Eq.~\eqref{eq:density_states} with a Gaussian factor,
\begin{align}
    W(E,\varepsilon_n^{l},\Gamma) = \frac{1}{\sqrt{2\pi \Gamma^2}}e^{-(E-\varepsilon_n^{l})^2/(2 \Gamma^2)}.
\end{align}
The value for $\Gamma$ may be determined from considerations of disorder in a real system. We adopt for it the value typically inferred from the decay of SdH oscillations, referred to as the Dingle temperature \cite{Shoenberg1984}.
Thus we define the electron density at $E_F$ via,
\begin{align}
    n(E_F,B) = \frac{1}{2\pi l_B^2} \sum_{n,l}\int_{-\infty}^{E_F} dE W(E,\varepsilon_{n}^{l})f(\varepsilon_{n}^{l},E_F).
    \label{eq:fixed_density}
\end{align}
Here we introduced the Fermi Dirac occupation factor $f(E,\mu)$. Clearly, for fixed density, Eq.~\eqref{eq:fixed_density} must be solved self-consistently. To do so, we fix the density, and integrate the r.h.s of Eq.~\eqref{eq:fixed_density} until we converge to the set value of $n$.
The Hall conductivity is defined through the Kubo formula \cite{Ando1974_1,ando1974_4}. At $E_F$, we have,
\begin{align}
    \notag \sigma^{xy} = \frac{i \hbar e^2}{\pi l_B^2}\sum_{n,l,n',l'} &\frac{(f(\varepsilon_n^{l},E_F) - f(\varepsilon_{n'}^{l'},E_F)) }{(\varepsilon_n^{l} - \varepsilon_{n'}^{l'})^2} \times \\ & 
    \langle n,l | v^{x} | n',l' \rangle \langle n',l' | v^{y} | n,l \rangle. 
    \label{eq:hall}
\end{align}
Here, the velocity operators are given by the derivative of the Hamiltonian with respect to the kinetic momentum. Letting $\boldsymbol{\pi} = \mathbf{p} + e\mathbf{A}$,
\begin{align}
    v_x = \frac{\partial H}{\partial \pi_x} =  \frac{p_x}{m^{*}} \boldsymbol{1}+\frac{\alpha}{\hbar}\sigma_y, \\
    v_y = \frac{\partial H}{\partial \pi_y} =  \frac{p_y}{m^{*}} \boldsymbol{1}-\frac{\alpha}{\hbar}\sigma_x.
    \label{eq:ops}
\end{align}
For the bilayer case, the Pauli matrices $\sigma$ are additionally dressed with $\tau_z$. We stress that $v_x, v_y$ do not have any diagonal components, reflecting the incompressible nature of the quantum Hall fluid, even in the presence of SOC \cite{WangVasilopoulos2005,winkler2003spin}.
As we do not solve for the exact wavefunctions in the presence of disorder, we approximate Eq.~\eqref{eq:hall} with the formula $\sigma^{xy} = \frac{e^2}{h} \sum_{n,l}f(\varepsilon_{n}^l)$. This approximation is justified in the limit where the disorder strength is such that delocalized states are situated only at the center of the Hall plateau. It is also the zero temperature, zero disorder limit of Eq.~\eqref{eq:hall}.
\subsection{Single band magnetotransport with SOC}
\begin{figure}[ht]
    \centering
    \includegraphics[width=\columnwidth]{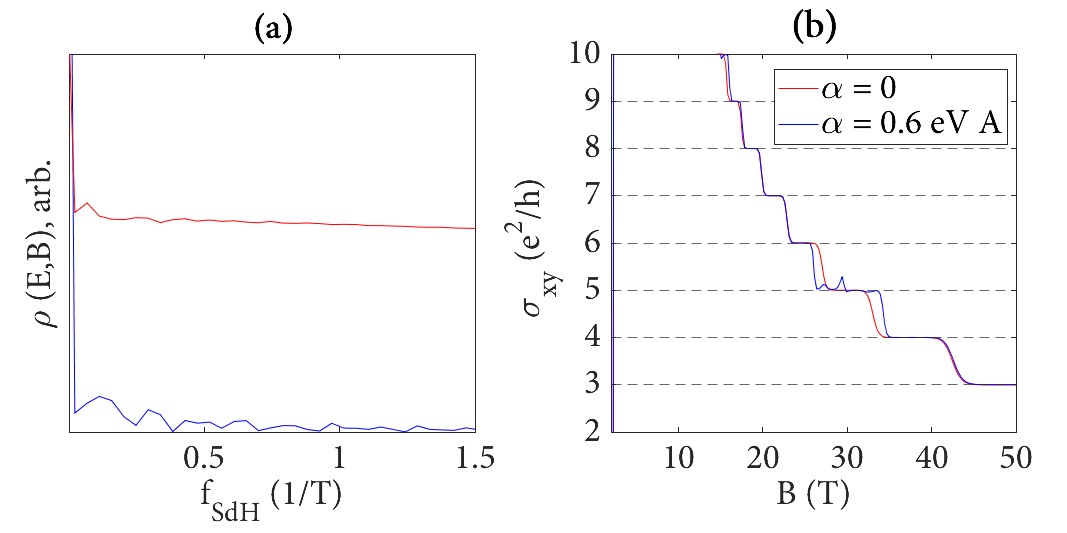}
    \caption{ 
    SdH oscillations in the density the Hall conductivity for a single band with SOC. (a) For $\alpha = 0$, a single peak around $f \approx 0.005 T^{-1}$ is observed, which splits when $\alpha \neq 0$. The red curve was shifted by constant value. This is the signature of the split of the Fermi surface into two pockets with opposite spin texture. (b) Hall conductivity as a function of $B$ for fixed density. Note that the only odd plateaus are observed. This is consistent, as even at $\alpha = 0$, Landau levels are non-degenerate due to Zeeman splitting. For $\alpha \neq 0$, no significant change is observed (besides a minor shift of the plateaus), as the Hall conductivity is quantized regardless of SOC. The density is fixed at $n = 3.5 \cdot 10^{12} \mathrm{cm}^{-2}$. The broadening is taken to be $\Gamma = 10.5 K$.}
    \label{fig:fig7}
\end{figure}

For the case of the single band with SOC, we solve Eq.~\eqref{eq:fixed_density} for a fixed density for different values of $\alpha$, and $B$. We then calculate the Hall conductivity. We Fourier transform $n(E)$ as a function of $B$ in units of $B^{-1}$. We label this quantity $\rho(B)$. In Fig.~\ref{fig:fig7}(a) we plot the frequencies of the oscillations of $\rho(B)$, which show a single peak (with higher harmonics) for $\alpha = 0$; this peak splits into two separate frequencies for $\alpha \neq 0$. This is consistent with the dispersion plotted in Fig.~\ref{fig:main_fig2}(a), which shows the formation of two pockets of opposite spin texture. 
The Hall conductivity, plotted in Fig.~\ref{fig:fig7}(b) does not show any appreciable difference between $\alpha = 0, \alpha \neq 0$. This is due to the existence of Zeeman splitting even at $\alpha = 0$, already lifting the Landau level degeneracy even in the absence of SOC. This result complements in Ref.~\cite{WangVasilopoulos2005}, whose results are recovered by setting $g_0 = 0$.
\subsection{Magnetotransport with hidden spin polarization}
\begin{figure}[ht]
    \centering
    \includegraphics[width=\columnwidth]{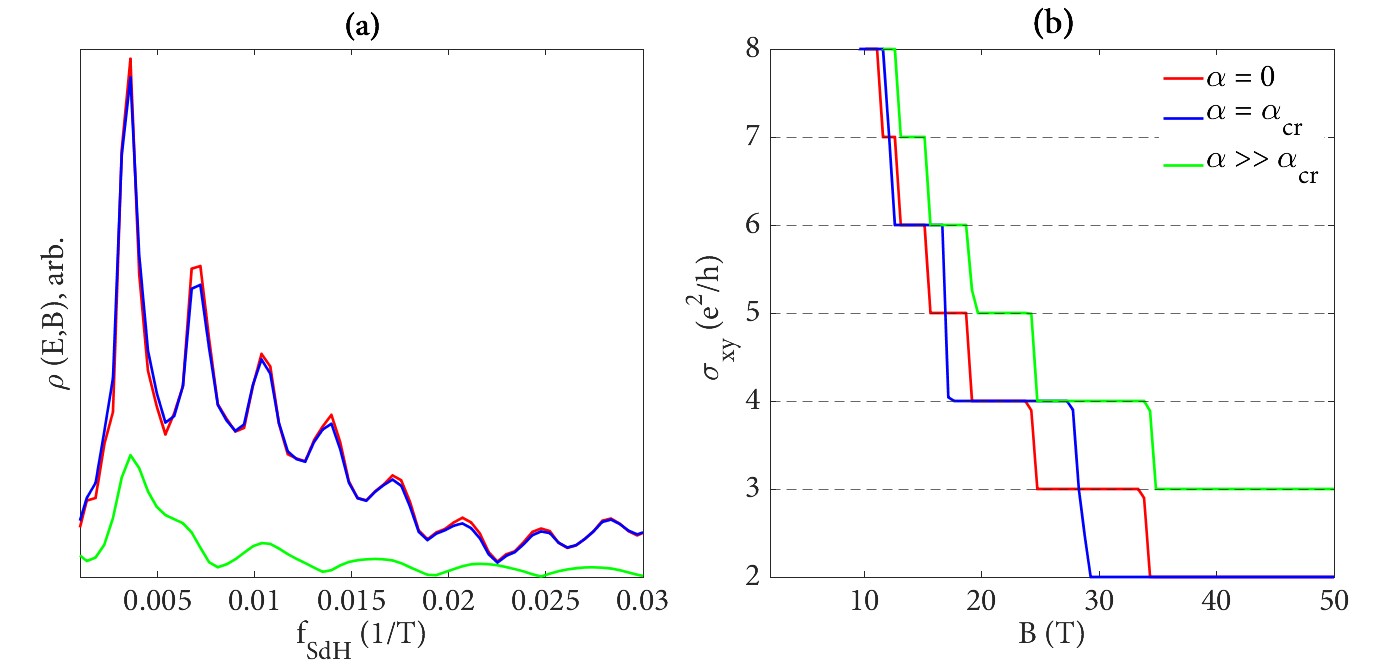}
    \caption{Quantum oscillations and Hall conductivities for the model with hidden spin polarization. (a) SdH oscillations in density as a function of $1/B$. No change is observed for $\alpha = 0$ and $\alpha \sim \alpha_{\mathrm{cr}}$, consistent with the band structures shown in Fig.~\ref{fig:fig4}(a)-(b), where only a single Fermi surface is apparent. (b) Hall conductivity. For $\alpha = 0$, both even and odd plateaus are apparent. For $\alpha \to \alpha_\mathrm{cr}$ only even plateaus appears. For $\alpha \gg \alpha_{\mathrm{cr}}$, odd plateaus are recovered. Here, $\Gamma = 10.5K, t=0.4\mathrm{eV}$, $n = 2 \cdot 10^{12} \mathrm{cm}^{-2}$, $m^{*} = 0.2 m_0$.}
    \label{fig:fig8}
\end{figure}

\begin{figure}[ht]
    \centering
    \includegraphics[width=\columnwidth]{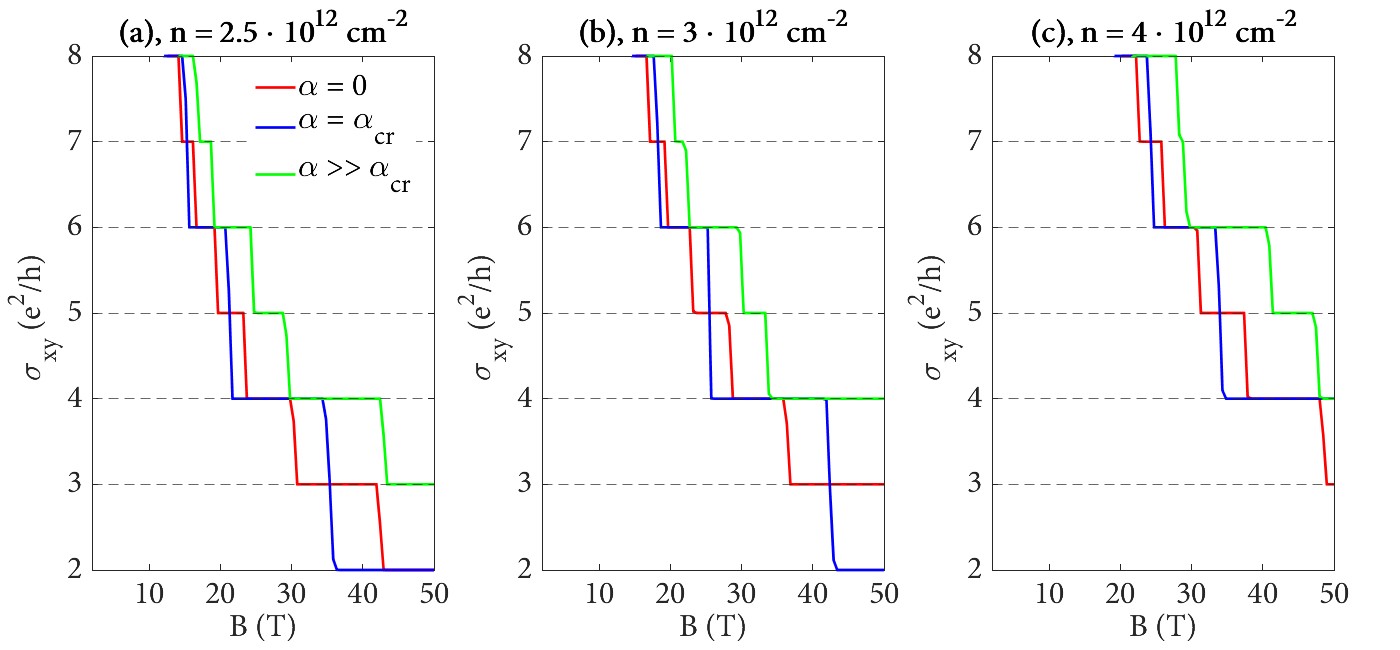}
    \caption{Hall conductivities for variable densities. Suppression of odd Hall plateaus at $\alpha_{\mathrm{cr}}$ apparent for a wide range of values. Parameters identical with Fig.~\ref{fig:fig8}.}
    \label{fig:fig9}
\end{figure}

For the bipartite case, we take the velocity operators in Eq.~\eqref{eq:ops} dressed with $\tau_z$ for the layer index. We once again fix the density and calculate the quantum oscillations and Hall conductivity for several densities. We then tune the SOC strength to match the conditions seen in Fig.~\ref{fig:fig6}(a). For $\alpha \approx \alpha_\mathrm{cr}$, derived from Eq.~\eqref{eq:main1}, we find a suppression of all odd plateaus in the Hall conductivity up to a representative large magnetic field $B=50T$. This is demonstrated in Fig.~\ref{fig:fig8}(b). This behavior persists for a range of densities, as we show in Fig.~\ref{fig:fig9}.
In quantum oscillations however, no dramatic change is observed until $\alpha \gg \alpha_{\mathrm{cr}}$, where two Fermi surfaces appear, as shown in Fig.~\ref{fig:fig4}(c). That the quantum oscillations retain only one frequency, as seen in Fig.~\ref{fig:fig8}(a) distinguishes this regime of magnetrotransport from those in materials without hidden spin texture. The enforcement of a single Fermi surface, at low densities, is a direct consequence of inversion symmetry in the bulk. 
That even plateaus, in agreement with the picture of an induced degeneracy of Landau levels due to the compensation of SOC by inversion symmetry, accompanied by a \textit{single} Fermi surface is the main result of this work.
\subsection{Role of inversion symmetry}
\begin{figure}[ht]
    \centering
\includegraphics[width=\columnwidth]{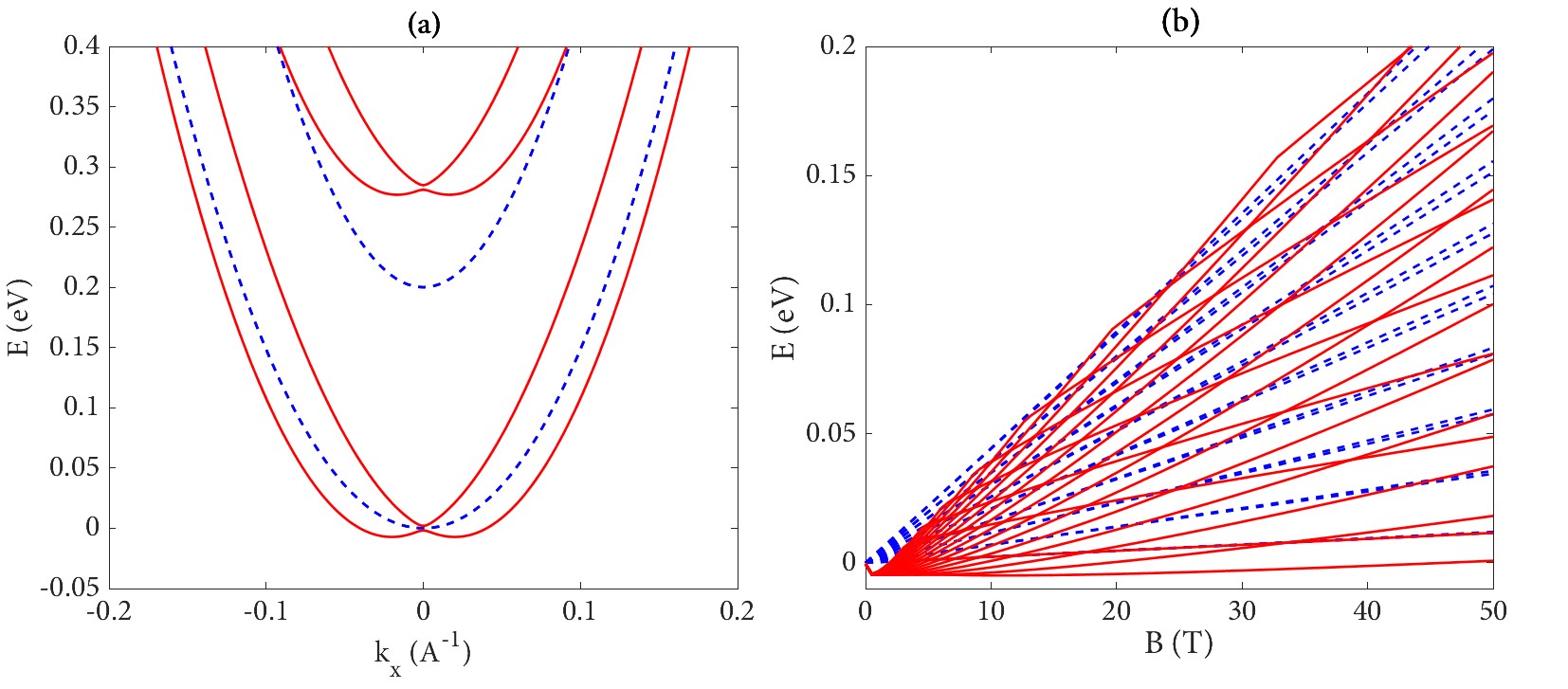}
    \caption{Dispersion and Landau levels of an inversion broken system. (a) Band structure, with doubly degeneracy lifted, exhibiting two Fermi surfaces with inversion-odd spin texture. (b) Landau levels of the inversion broken system. Multiple Landau level crossings appear, with the splitting substantially enhanced compared to $E_z$. (Dashed blue lines are the values for the inversion symmetric case $\Delta = 0$). Here, $t = 0.1$, $m^{*} = 0.2 m_0$, $\Delta = 0.1 \mathrm{eV}$, $\alpha = \alpha_{\mathrm{cr}}$. }
    \label{fig:fig10}
\end{figure}

\begin{figure*} 
\centering
\includegraphics[width=0.95\textwidth]{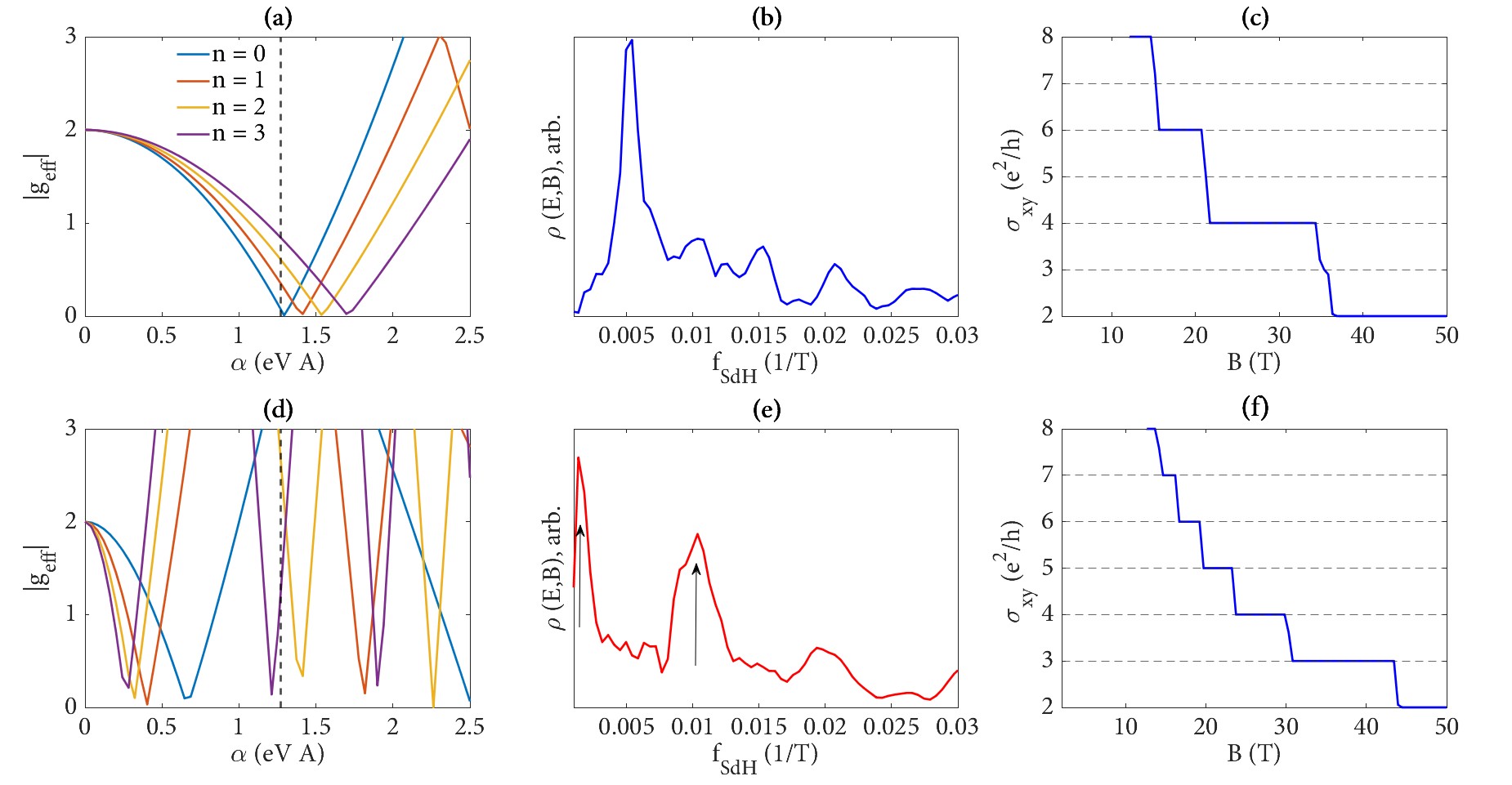}
    \caption{Inversion symmetric and inversion symmetry-broken magnetotransport.
    (a,d) Effective $g$ factor for the inversion symmetric and broken cases, respectively. Dashed line shows $\alpha_{\mathrm{cr}}$. (b,e) SdH oscillations with/without inversion symmetry. The single peak in (b) is split into two peaks (e), as denoted by black arrows. (c,f) Hall conductivity showing even plateaus for inversion symmetry (c), and both even/odd plateaus when the symmetry is broken in (f). Here, $\Gamma = 10.5K$, $m =0.4m_0$, $t=0.1 \mathrm{eV}$, $\Delta = 0.1 \mathrm{eV}$, $\alpha = \alpha_\mathrm{cr}$, $n = 2.5 \cdot 10^{12} \mathrm{cm}^{-2}$. For panels $(a),(d)$, $B = 50 \mathrm{T}$.}
    \label{fig:fig11}
\end{figure*}

We now show that the degeneracy of Landau levels at order $t^{-1}$ at $\alpha = \alpha_{\textrm{cr}}$ is protected by inversion symmetry. To account for inversion symmetry breaking in the model, we recall that the suppression of the effects of SOC is possible due to the identical nature of the two layers/sublattices. We introduce an asymmetry strength between the two sub-units, $\Delta$. The Hamiltonian with the asymmetry reads,
\begin{align}
    H = \frac{p^2}{2m^{*}}\boldsymbol{1} + \frac{\alpha \tau_z}{\hbar}\left(\sigma_y p_x - p_y\sigma_x\right) + t \tau_x + \Delta \tau_z.
\end{align}
This Hamiltonian no longer commutes with $\mathcal{P}$ and the double degeneracy is lifted. We define a renormalizedhybridization strength  $\tilde{t} = \sqrt{t^2 + \Delta^2}$ and the degree of a symmetry by $\varphi =\tan^{-1}(\Delta/t)$. Now, the inter-sublattice and onsite terms are $\tilde{t}(\cos(\varphi),0,\sin(\varphi)) \cdot \vec{\tau}$. The spectrum is given by,
\begin{align}
    \varepsilon_{n,\sigma,\tau} = \frac{p^2}{2m^{*}} + \sigma\sqrt{\tilde{t}^2+\alpha^2 p^2/\hbar^2+2\tau\alpha |p|\tilde{t} \sin(\varphi)/\hbar}.
\end{align}
The breaking of inversion symmetry allows for a finite expectation value for $\langle \tau_z \rangle$ at any momentum, and thus a finite spin texture. The dispersion is plotted in Fig.~\ref{fig:fig10}(a).
To derive the Landau spectrum, we note that since $\Delta \tau_z$ commutes with the SOC term, the problem can be again diagonalized exactly with the inversion breaking term. To leading order in $\Delta$, we find the corrections to the energy splitting is, for the branch $\tau=-1$,
\begin{align}
    \notag &\varepsilon_{n,+,\tau=-1} -\varepsilon_{n,-,\tau=-1}  = \\ \notag & E_z -\frac{\alpha^2}{l_B^2 t} \biggl(1 + 
    \frac{2(2n+1)\Delta^2}{t |\hbar \omega_c - E_z|} -\frac{(2n+1) |\hbar \omega_c - E_z|}{2t}\biggr) \\ & + \mathcal{O}(t^3, \Delta^4).
\end{align}
The associated Landau spectrum is shown in Fig.~\ref{fig:fig10}(b).
We now present the consequences for magnetotransport for inversion symmetry breaking. In Fig.~\ref{fig:fig11}(a,d) we plot the difference in $g_\textrm{eff}$, with and without inversion symmetry breaking.
Clearly, at the unperturbed value for the critical $\alpha$, when $\Delta \neq 0$, the $g$ factors are enhanced, distributed unevenly for all Landau levels (c.f. Fig.~\ref{fig:fig6}(a)-(b)) lifting all remaining degeneracies. As a result, we observe splitting of the main frequency in quantum oscillations (Fig.~\ref{fig:fig11})(b,e)) and finally, the emergence of odd Hall conductivity plateaus (Fig.~\ref{fig:fig11})(c,f)). 
\section{Multi-layer case}
\label{sec:multilayer}
We discuss the extension of our results to the case of a multi-layer/multicomponent system, relevant for heterostructures with tunable symmetry properties. We start by designating the inversion-symmetric elemental unit as the \textit{unit cell} for the multilayer problem. Details of the model and the band structure of the multilayer system are presented in App.~\ref{app:multilayer}.
We outline the ingredients for the multicomponent model: a unit cell (Fig.~\ref{fig:fig12}(a)) is coupled to neighboring cells with an additional parameter $t'$. The resultant band structures, shown in Fig.~\ref{fig:fig12}(b)-(d) present the sub-bands that arise due to the inter-unit cell coupling. 
As the coupling between unit cells connect the top layer of unit cell $N$ to the bottom layer of unitcell $N+1$, the Hamiltonian of the multi-cell system has the form,
\begin{align}
    H = \left(\begin{matrix} \\ 
    H^{N-1}_{2 \times 2}(-\alpha) & t'\sigma_0 & 0 & 0 \\
    t'\sigma_0 & H_{2 \times 2}^{N} (\alpha) & t \sigma_0 & 0 \\
    0 & t\sigma_0 & H_{2 \times 2}^{N} (-\alpha) & t'\sigma_0 \\
    0 & 0 & t'\sigma_0 & H_{2 \times 2}^{N+1} (\alpha),
    \end{matrix}   
    \right)
    \label{eq:Hmulti}
\end{align}
Where $H_{2\times 2}$ is the single-band Hamiltonian in Eq.~\eqref{eq:main_singleband_ham}, and $\sigma_0$ is the identity matrix in the spin basis.
The quantization with a magnetic field of Eq.~\eqref{eq:Hmulti} proceeds with the substitution of minimally coupled momenta in the appropriate two-by-two block with the appropriate SOC strength $\pm \alpha$. An analytical solution is impracticable for a Hamiltonian of these dimensions. In what follows we diagonalize the Hamiltonian numerically with $M \gg 1$ LLs included in the basis specification (the Hamiltonian has dimensions $(N_{uc} \times M \times 4)^2$. We have checked that $M$ is large enough, such that the energies of the lowest branch are fully converged. To make the multilayer case concrete we present a material example which involves a multilayer model. 
\subsection{Material candidates}
\label{sec:matexam}
Bi\textsubscript{2}O\textsubscript{2}Se is rapidly emerging as a semiconductor with highly advantageous properties, with high carrier mobility and a moderate band gap \cite{wu2017high,Li2021_review_BOS,ding2022bi2o2se}. Moreover, the Bi content of the material suggests the strong role that SOC plays in the system. The bulk is arranged from layers of Bi\textsubscript{2}O\textsubscript{2} separated by Se in a ``zipper-like" mechanism \cite{wei2019quasi}. Both the bulk \cite{wu2017high} and the multilayered structure arrange in the space group I4/mmm which includes the inversion operation, making this material ideal for probing the interplay of strong SOC, inversion symmetry, and magnetotransport. We performed first-principle calculations on Bi\textsubscript{2}O\textsubscript{2}Se multilayer structures. The details of the calculation are presented in App.~\ref{app:materialcand}. 
For a 4 unit cell system, we find excellent agreement with the model of Eq.~\eqref{eq:Hmulti} with the parameters $t=t' = 0.2\mathrm{eV}$ and $m^{*} = 0.14m_0$, $\alpha = 1.45 \mathrm{eV \AA}$. 
For this system, $\alpha_{\textrm{cr}}/g_0 = 1.38$, i.e., very close to the ideal value of $\alpha_{\textrm{cr}}/g_0 = 1$. The suppression of odd Hall plateaus is presented in App.~\ref{app:materialcand}, Fig.~\ref{fig:fig13}. 
Other candidates include the recently discovered Si\textsubscript{2}Bi\textsubscript{2} \cite{lee2020unveiling} and InTe\cite{lee2023unconventional}, both with exceptionally large on-site SOC, making them suitable via Eq.~\eqref{eq:main1}.
\section{Discussion}
In this work, we presented a new magnetotransport feature inherent to systems with hidden spin polarization. We started with a review of magnetic field induced dynamics in a single-band with SOC, finding that the main effect of SOC was the splitting of Fermi surfaces and the lifting of any remaining (non-accidental) degeneracies in the electronic spectra. Here, the effect of SOC is manifest due to the explicit breaking of inversion symmetry in the model. Next, we generalized the problem to explicitly include inversion symmetry while maintaining SOC. The inversion symmetry was enforced by adding another degree of freedom -- layer/sublattice -- in which the strength of SOC was flipped such that the \textit{overall} structure retained inversion symmetry, in keeping with the picture of hidden spin polarization introduced in Ref.~\cite{zhang2014hidden}. We then showed how the dynamics in this newly-formed unit cell differ substantially from the single-band case. Our main findings were the dramatic suppression of Landau level splittings as a function of SOC for a wide range of densities. As a result, two distinctive features are shown in magnetotransport: (1) the existence of \textit{even-integer} steps in the Hall conductivity of the system; (2) the observation of a \textit{single} peak in quantum oscillations -- meaning that the unique quantum Hall effect in this system is \textit{not} driven by multiple electron pockets that trivially double the quasiparticle content of the system. 

Finally, we showed how the basic inversion symmetry element -- the unit cell -- can be stacked further to include many layers, further showing that this model accounts well for the low energy properties of a layered semiconductor, Bi\textsubscript{2}O\textsubscript{2}Se. Moreover, the strength of SOC in the system was such that for typical parameters of LL broadening, the robustness of the even-integer Hall plateaus is evident and should be measurable across a variety of electronic densities, as was recently confirmed experimentally in Ref.~\cite{Wang2024}.
\paragraph*{Landau level splitting---} 
The chief quantity governing the emergence of even-integer plateaus is the Landau level splitting. From this,  we calculated the effective $g$-factor, introduced in Eq.~\eqref{eq:main_gfactor}. We showed how the $g$ factor evolves as a function of SOC in a system with and without inversion symmetry. For the single-band case, SOC always leads to a strong enhancement of the $g$ factor, culminating in substantial LL mixing. This is due to the fact that in this constrained Landau problem, for large $\alpha$, $\alpha l_B$ is always the dominant energy scale. Competition with Zeeman splitting leads to a sign-reversal of the $g$ factor, but this effect is strongly Landau-level index dependent. For weak fields (but finite SOC), the effective splitting always \textit{increases} as a function of decreasing field. Famously, this explains the observation of finite spin splitting for small $B$ in InGaAs \cite{Das1989}, where the effect would otherwise be small from the atomic/lattice point of view. When the problem is extended to the inversion-symmetric case, a different behavior emerges. SOC cannot directly compete with Zeeman splitting, as the state when $t \to \infty$ prevents any finite expectation value for the SOC at the leading order. Instead, corrections to the Zeeman splitting enter quadratically, and involve \textit{higher} (in energy) branches of the dispersion. In this situation, SOC can negate the effect of Zeeman splitting, and because of the nature of SOC, can do so \textit{independently} of the LL index.  This leads to the suppression of the Zeeman splitting. Furthermore, suppression of the spin splitting, therefore, extends to $B \to 0$, unlike the single band case.
\paragraph*{Measuring the effective $g$ factor---}
The finding of even integer plateaus in the conductivity relies on the suppression of the spin-splitting of Landau levels. For finite $t$, the splitting for LLs becomes LL index dependent, with the correction suppressed at order $t^{-2}$. The question of whether the residual splitting is picked up by experimental measurements depends on magnetic field strength (determining $\Delta \varepsilon$), electron density and proximity of $\alpha$ to the critical value. In addition, the LL broadening parameter $\Gamma$ plays a crucial role, effectively smearing the contribution of LLs whose energy difference $\Delta \varepsilon \lesssim 2\Gamma$. In experiments, it is possible to determine the effective splitting and bound $g_\textrm{eff}$. Assuming the minimal longitudinal conductivity $\sigma_{xx}$ to have an activated behavior $\sigma_{xx} \sim e^{-\Delta \varepsilon/k_B T}$ (where $T$ is the experiment temperature), LLs will be observed if $\Delta \varepsilon \ll k_B T$. But the broadening of LLs ($\Gamma$) would lead to the condition that $\Delta \varepsilon - \Gamma < 2 k_B T$ \cite{bandurin2017high}. For a typical experimental temperature of $T = 4\mathrm{K}$, at $B = 50\mathrm{T}$ and $\Gamma = \pi k_B T_D$ \cite{Shoenberg1984}, we find that $g_\mathrm{eff} < \textrm{1.27}$, for a Dingle temperature of $T_D \sim 11\mathrm{K}$. As a result, a considerable number of LL splittings will remain suppressed; this is seen in Fig.~\ref{fig:fig11}(a), where $\alpha = \alpha_{\mathrm{cr}}$ the first 4 LL splittings fall below the experimental threshold of $g \sim \textrm{1.27}$.
\paragraph*{Inversion symmetry breaking and effective SOC---}
By breaking inversion symmetry, the normal quantum Hall state with both even and odd Hall plateaus is recovered. Two approaches can be applied to break inversion symmetry: an out of plane electric field and the proximatizing of the system to a substrate. In the latter case, we point out that the effect would be most appreciable for the electronic states localized close to the surface. In the case of the multilayer (Sec.~\ref{sec:multilayer}) system we note that the effect of the substrate depends on the wavefunction of the lowest lying states. Assuming confinement from the sample edge, the wavefunction of the ground state is expected to behave as $|\psi(z \to 0)|^2 \sim L^{-1}$ and is strongly suppressed for thicker systems. Consequently, the degree of inversion symmetry breaking -- and the splitting of Landau levels -- can be tuned by controlling the number of layers and thickness of the sample.
\paragraph*{SOC mechanisms---}
The foregoing results were presented for the case of hidden Rashba SOC. We now comment on the situation with different spin-orbit mechanisms. From the point of the hidden Dresselhaus effect, similar considerations as we applied in Sec.~\ref{sec:spintext} are relevant. Using perturbation theory, the leading order correction for the ground state energy remains negative, provided the states that are coupled exchange LL and spin indices, which is true for the Dresselhaus effect as well \cite{winkler2003spin}. One key distinction however, must be made for Ising SOC. In this case, we note that for any momentum $[H, \sigma_z] = 0$, even in the presence of $B$. This rules out the possibility of SOC competing with Zeeman splitting. Put more simply, first write the bipartite Hamiltonian with Ising-like SOC but without Zeeman splitting. This is:
\begin{align}
    \notag H &= \hbar\omega_c(n+1/2) \delta_{nm} \boldsymbol{1} +  + t \tau_x \delta_{nm} + \\ & \frac{\alpha}{l_B} \sigma_z \tau_z \left(\sqrt{m+1}\delta_{n,m+1} + \sqrt{m}\delta_{n,m-1}\right).
\end{align}
It is trivial to see that this Hamiltonian commutes separately with $\mathcal{P} \mathcal{T}$ and $\sigma_z$. Without Zeeman splitting, all eigenstates are doubly degenerate because of $\mathcal{P} \mathcal{T}$. However, since the states are now diagonal in $\sigma_z$, the addition of $H_z = \frac{1}{2} g_0\mu_B B$ will trivially couple to the eigenstates of $\sigma_z$ leading to constant splitting of $\Delta \varepsilon = E_z$, independent of $\alpha$. Thus, there will be no suppression of the LL splitting, which will remain the bare $g$. 
The dispersion however remains unchanged, since $\varepsilon \sim \frac{p^2}{2m^{*}}\pm \sqrt{t^2+E_{\textrm{SOC}}^2}$. Therefore, the presence of a suppressed LL splitting is also direct evidence for a Rashba-like mechanism for the SOC in the studied system.  
\begin{acknowledgements}
We thank Erez Berg, Jingyue Wang, and Hailin Peng for illuminating discussions. D.K. is supported by the Abrahams Postdoctoral Fellowship of the Center for Materials Theory, Rutgers University, and the Zuckerman STEM fellowship. A.S. was supported by grants
from the ERC under the European Union’s Horizon 2020
research and innovation programme (Grant Agreements
LEGOTOP No. 788715 ) and the DFG (CRC/Transregio
183, EI 519/71). B.Y. acknowledges the financial support by the European Research Council (ERC Consolidator Grant ``NonlinearTopo'', No. 815869) and the ISF - Personal Research Grant	(No. 2932/21) and the DFG (CRC 183, A02).

\end{acknowledgements}
\bibliography{main.bbl}
\newpage
\clearpage

\appendix
\setcounter{equation}{0}\renewcommand\theequation{A\arabic{equation}}
\section{Single band Landau quantization with SOC}
\label{app:singleband}
In this section, we review Landau quantization in the canonical single band model, introduced in Eq.~\eqref{eq:main_singleband_ham}, with SOC.
In Eq.~\eqref{eq:single_band_H} we presented the spinful single band Hamiltonian. We follow previous works \cite{Schliemann2003,WangVasilopoulos2005,Zhang_2006} in finding the exact solution for the dynamics with a magnetic field. We quantize the problem with a magnetic field by choosing the Landau gauge such that $\mathbf{p} \to \mathbf{p} +  e\mathbf{A}$. Applying a perpendicular magnetic field $\mathbf{B} = B \hat{z}$, we use the gauge $\mathbf{A} = (0, Bx, 0)$. We additionally introduce the Zeeman spin coupling $H_z = \frac{g \mu_B B}{2} \sigma_z$. The Hamiltonian reads,
\begin{align} 
\notag
    H &= \frac{1}{2m^{*}} \left(p_x^2+(p_y+e B x)^2\right) + \\ & \frac{\alpha}{\hbar} \left(\sigma_y p_x -\sigma_x (p_y+e B x) \right) + \frac{g \mu_B B}{2} \sigma_z.
    \label{eq:singleband_ham}
\end{align}
We first note that $[H,p_y] = 0$ allowing us to replace $p_y$ with its expectation value $p_y = \hbar k_y$. Then, we define the shifted coordinate center, $x' = x_0 + x$, with $x_0 = \frac{\hbar k_y}{eB}$. We switch to the ladder operator description of Landau levels, by the substitution $p_{x'} = \frac{i\hbar}{\sqrt{2} l_B} \left(a^\dagger - a\right)$, and $x' = \frac{l_B}{\sqrt{2}} (a + a^\dagger)$, such that $[a,a^\dagger] = 1$, and $[x', p_{x'}] = i \hbar$. Here, $l_B = \sqrt{\frac{\hbar}{eB}}$ is the magnetic length. Inserting the operators into Eq.~\eqref{eq:singleband_ham} one obtains, 
\begin{align}
\notag
    H &= \hbar \omega_c \left(a^\dagger a + \frac{1}{2}\right) - \\ & \frac{\sqrt{2}\alpha}{l_B}\left(a \sigma^{+}+ a^\dagger \sigma^{-}\right) + \frac{g \mu_B B}{2} \sigma_z,
    \label{eq:ham_in_basis}
\end{align}
where $\sigma_{\pm} = \frac{1}{2}\left(\sigma_x \pm i\sigma_y\right)$ and the cyclotron frequency is $\omega_c = \frac{e B}{m^{*}}$. The eigenstates of Eq.~\eqref{eq:ham_in_basis} are obtained by projecting the Hamiltonian on Landau levels with definite level index $n$ and $\sigma = \uparrow, \downarrow$. Denoting these states as $\left|n, \sigma\right\rangle$, the matrix elements $\langle n, \sigma | H | m, \sigma' \rangle$ are  
\begin{align}
\notag 
&H_{nm}^{\sigma \sigma'} = \left[\hbar \omega_c \left(n + \frac{1}{2}\right) + \frac{\sigma' g \mu_B B}{2} \right]\delta_{n,m} \delta_{\sigma, \sigma'} - \\ & \frac{\sqrt{2} \alpha}{l_B} \left(\sqrt{m} \delta_{n,m-1} \delta_{\sigma,\uparrow}\delta_{\sigma' ,\downarrow}+\sqrt{m+1}\delta_{n,m+1} \delta_{\sigma,\downarrow}\delta_{\sigma' ,\uparrow}\right).
\label{eq:fullham}
\end{align}
In this form, the Hamiltonian assumes a nearly tri-diagonal form which can be diagonalized exactly. The solution for finite $\alpha$ and $g$ mixes both spinor states and original Landau level indices. The eigenvalues are given in terms of an integer $n$ and renormalized spin state $\tilde{\sigma} = \pm 1$,
\begin{align}
    \notag \varepsilon_{n, \sigma= \pm 1} &= \left(n+\frac{1}{2}(1+\sigma)\right)\hbar \omega_c -\\ &  \sigma \sqrt{\frac{2(n+(1+\sigma)/2)\alpha^2}{l_B^2} + \left(\frac{E_z}{2}\right)^2\left(1-\frac{2 m_0}{g m^{*}}\right)^2}.
    \label{eq:spectrum_singleband}
\end{align}
For compactness, we denoted $E_z = g \mu_B B$, which is the Zeeman energy. To understand this result, we look at the limit of vanishing SOC $\alpha \to 0$. The natural expectation, as seen from an immediate diagonalization of Eq.~\eqref{eq:fullham}. Using the fact that $\frac{2 m_0 E_z}{g m^{*}} = \hbar\omega_c$, for $\alpha = 0$, the spectrum reduces to $\varepsilon_{\pm} = \hbar\omega_c (n+1/2) \pm E_z$. In the limit of large magnetic field (that is $g \mu_B B \gg \alpha l_B^{-1}$), the spectrum of equally spaced Landau levels is again separated by the Zeeman term, while in the opposite limit (where SOC dominates), the spectrum takes the form $\varepsilon_{n,\pm} \approx n\hbar \omega_c \pm \sqrt{2n} \alpha l_B^{-1}$. For large $n \gg 1$, semiclassical arguments \cite{winkler2003spin,Li2016} dictate that the Landau levels have the form $\varepsilon_{n,\pm} \approx E_F \pm \alpha k_F$. This is due to the fact that the Landau level index (and consequently the density) is fixed by $E_F$ and $k_F \approx l_B^{-1}$ is enforced by the semiclassical quantization condition. 
\subsection{Level splitting}
The determination of the Landau level splitting establishes the effective $g$ factor and determines the linear response to a magnetic field. This quantity was defined in the main text in Eq.~\eqref{eq:main_gfactor}. To quantitatively explain the connection between Landau levels' splitting and SOC, we plot the effective $g$-factors as defined in Eq.~\eqref{eq:main_gfactor} with the relevant quantum numbers as a function of both $B$ and SOC strength $\alpha$.
\begin{figure}[ht]
    \centering
    \includegraphics[width=\columnwidth]{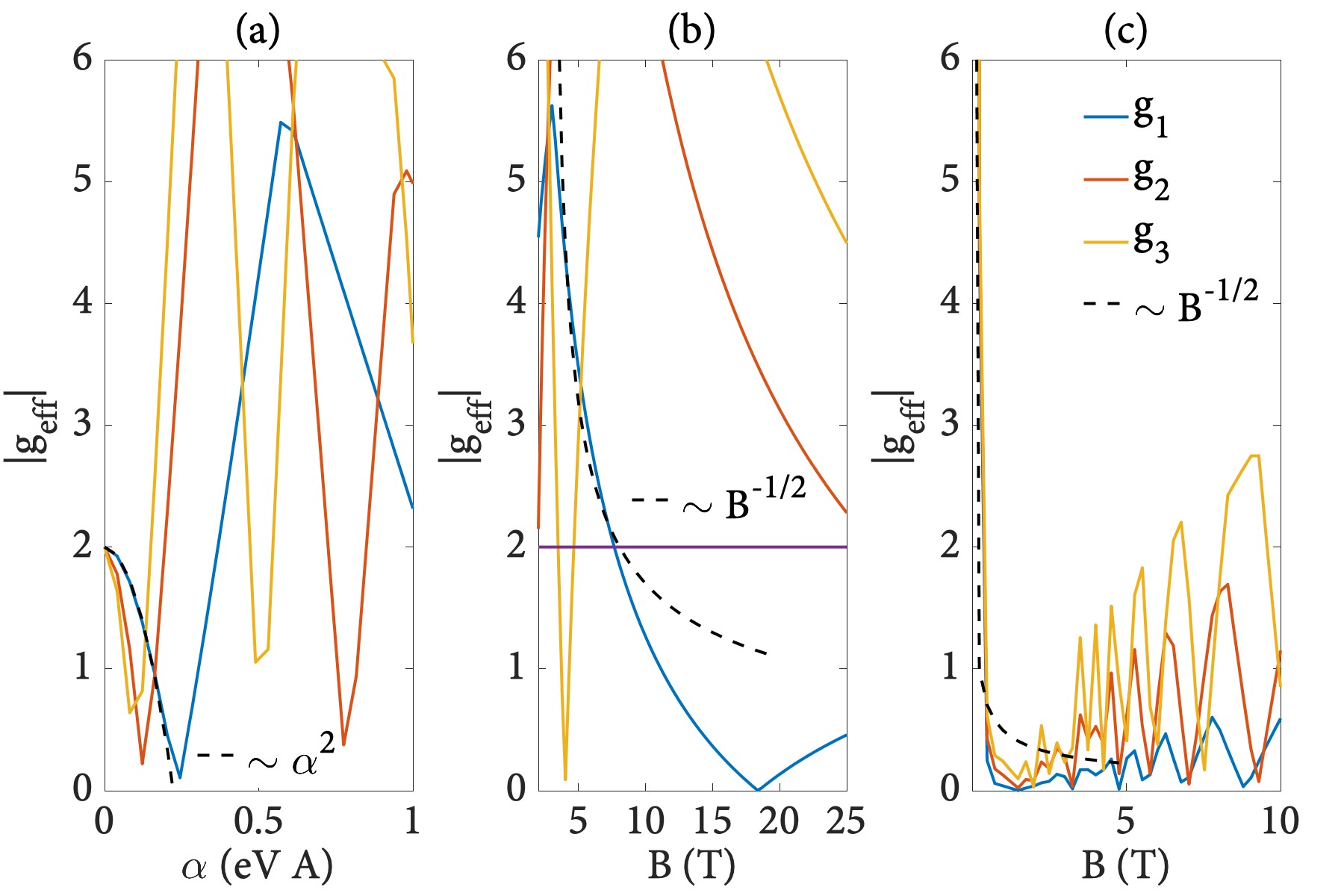}
    \caption{$g_{\mathrm{eff}}$ and Landau level splittings in a single electronic band with SOC. (a) $g_{\mathrm{eff}}$ as a function of $\alpha$ (SOC strength). Note the perturbative behavior of $g_{\mathrm{eff}} \sim \alpha^2$ for small $\alpha$. The jagged behavior is the result of Landau level crossings and the exchange of band indices. (b) Same as in, but as a function of $B$ for $\alpha = 0.2 \mathrm{eV} \AA$. For low fields, all $g_{\mathrm{eff}}$ diverge like $B^{-1/2}$. This realizes the limit of $\alpha l_B^{-1} \ll E_z$ (c) Same as in $(b)$ but for $\alpha l_B^{-1} \gg E_z$ ($\alpha = 1.2 \mathrm{eV}\AA$. Oscillatory behavior is observed. For all plots, colors refer to the splittings of Landau levels indexed by $n = 0,1,2$, respectively. In (b), the purple line corresponds to the splitting with $\alpha = 0$, which is simply equal to the bare Zeeman $g_0$.}
    \label{fig:fig3} 
\end{figure}
In the limit of $\alpha \to 0$, the Landau level splittings, denoted by $g_{n}$ -- $n$ being the LL index --  acquire the usual form and converge to the value $g_n/g_0 \to 1$. As $\alpha$ is increased, the $g$ factor decreases before changing sign at a critical value for $\alpha$ (note that we plot the absolute magnitude). This occurs roughly when $\alpha l_B^{-1} \approx g_0 \mu_B B$. The zero value for the splitting can be rationalized by the competition between in-plane and out-of-plane components of the spin projection. As SOC is increased, the $\sigma_z$ basis which defines the zero-SOC quantum numbers is no longer a faithful representation of the state. The spin state is rotated and is maximally mixed when $\alpha$ approaches the critical value. Importantly, the behavior in the vicinity of $\alpha \to 0$ is $\Delta \varepsilon \propto \alpha^2$ which comes from perturbative grounds, from states in which the spins are polarized along $\sigma_z$.
Beyond this regime, the Landau levels also cross as evidenced by many discontinuities in Fig.~\ref{fig:fig3}(a). After crossing, the trend is approximately $|g_{\textrm{eff}} |\propto |\alpha|$, as the eigenstates are of rotated spin states which allow for linear coupling to the SOC term in Eq.~\eqref{eq:ham_in_basis}. 
We presented the exact analytical solution and various limits in the main text. Here we point out that the solution is sensitively controlled by the parameter $q$. Whenever $q > 1$, the SOC will cause the effective $g$ to change sign; this is fulfilled for semiconductors whose quasiparticles have effective mass $m^{*} < m_0$ and $g_0 \approx 2$, as is commonly encountered for light electrons in semiconductor conduction bands. 
For the case $q=1$, Eq.~\eqref{eq:main_full_split} can be directly evaluated since $E_z$ disappears in its explicit form inside the expression. However, in this case $\hbar \omega_c = E_z$ and the recovery of expected behavior when $\alpha \to 0$ is easily verified. That the bare Zeeman splitting and SOC enter with different signs is confirmed numerically in the evaluation of $g$ factors for $n=0,1,2,\ldots$ Landau levels. Fig.~\ref{fig:fig3}(a) shows the initial quadratic decrease (for $q>1$) with increasing SOC at fixed $B$. At a critical value of $\alpha$, the effective $g$-factor is zero, whereupon it changes signs. It then increases roughly linearly with $\alpha$ due to Landau level and pseudospin mixing, which is expected when $\alpha l_B^{-1} \gg E_z$. Cusps and non-analytical behavior in $g_\textrm{eff}$ is seen for large values of $\alpha$ due to the intersection of Landau levels, as shown in Fig.~\ref{fig:fig3}(a)-(b). When Landau levels touch, they interchange their quantum numbers (since they are decoupled). As a result for fixed $n,\sigma$, the splitting changes slope and causes the sharp features in Fig.~\ref{fig:fig3}(b)-(c). Finally, the slope of the curves is proportional to the Landau level index $n$, indicating a non-universal value for which $g_\textrm{eff}$ turns zero. Eq.~\eqref{eq:main_perturb_level_splitting} shows that the value of $\alpha$ dictating the sign change is strongly $B$ dependent. This $B$ dependence reflects the underlying symmetry breaking in the band structure. For small magnetic fields, $g_{\textrm{eff}} \sim \frac{1}{\sqrt{B}}$ which explains the enhancement of the spin-splitting observed in semiconductor heterostructures, at zero field \cite{Das1989}. Finally, the appearance of the quantum limit is seen in the strong oscillatory behavior of $g_\mathrm{eff}$ as a function of $B$ in Fig.~\ref{fig:fig3}(c). This behavior is similar to the quantum oscillations observed in the density of states at fixed filling. 
\section{Multilayer systems}
\label{app:multilayer}
\begin{figure}[ht]
    \centering    \includegraphics[width=\columnwidth]{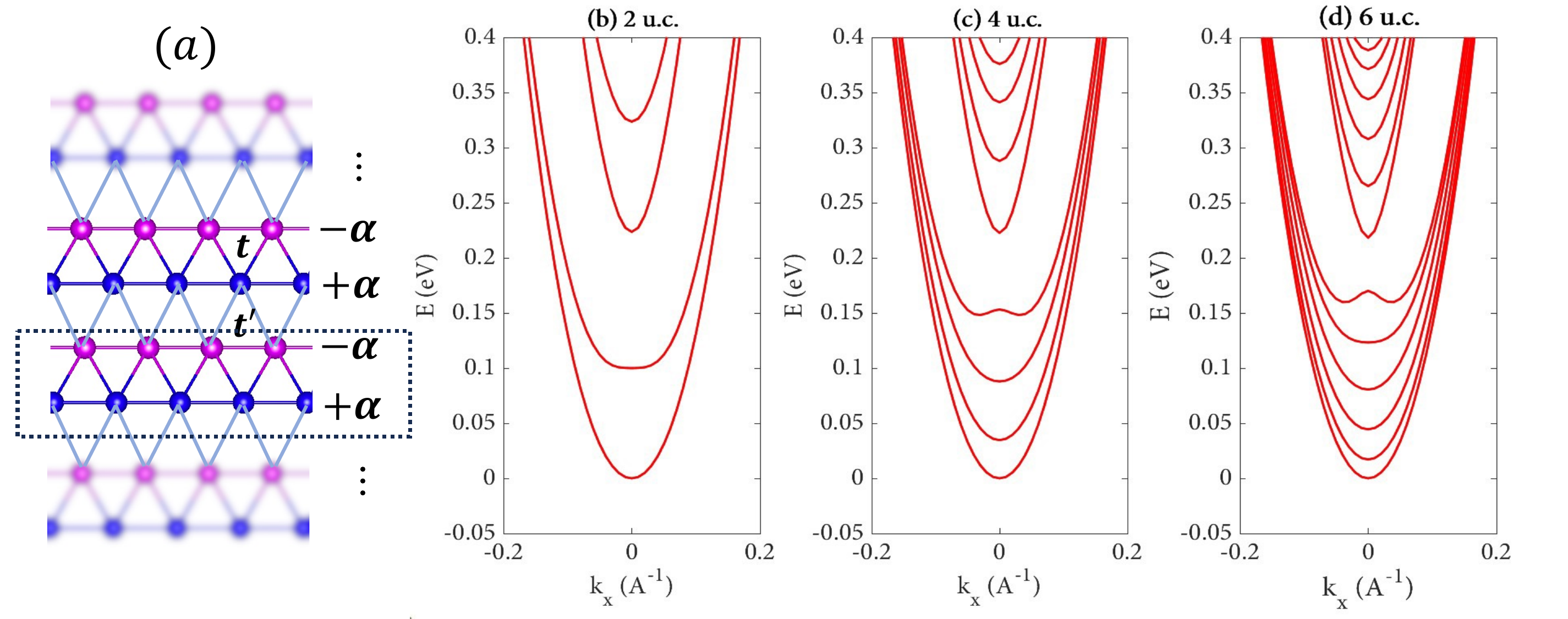}
    \caption{Multi-unit cell model and dispersions. (a) Construction of the multi-cell system: the inversion symmetric unit (marked with a dashed line) is coupled via a parameter $t'$ to neighboring cells. (b)-(d) Dispersion of the multicell model for 2, 4, 6 unit cells respectively, showing the subbands that result from the coupling between cells. Here $m = 0.2m_0$ and $\alpha = 2.0 \mathrm{eV \AA}, t= 0.2~\mathrm{eV}$.}
    \label{fig:fig12}
\end{figure}

In the main text, we introduced the Hamiltonian for a multi-unit cell system, as we presented in Sec.~\ref{sec:multilayer}. In Fig.~\ref{fig:fig12}(a) we present a sketch of stacked bilayer units with alternating atomic SOC with coupling between them. 
In Fig.~\ref{fig:fig12}(b)-(d) we illustrate how the renormalization of the band structure depends on the thickness -- i.e., the number of unit cells -- considered. Note that the lowest bands remain nearly parabolic as for these states, the wavefunction is mostly smoothly distributed across the entire slab, averaging over the alteranting SOC structure. For higher-lying states, whose density is centered on specific layers, the SOC cancellation is less pronounced leading to more significant renormalization. The number of sub-bands reflects the (increasing) thickness.
\section{First principle calculations on Bi\textsubscript{2}O\textsubscript{2}Se}
\label{app:materialcand}
\begin{figure}[t]
    \centering    \includegraphics[width=\columnwidth]{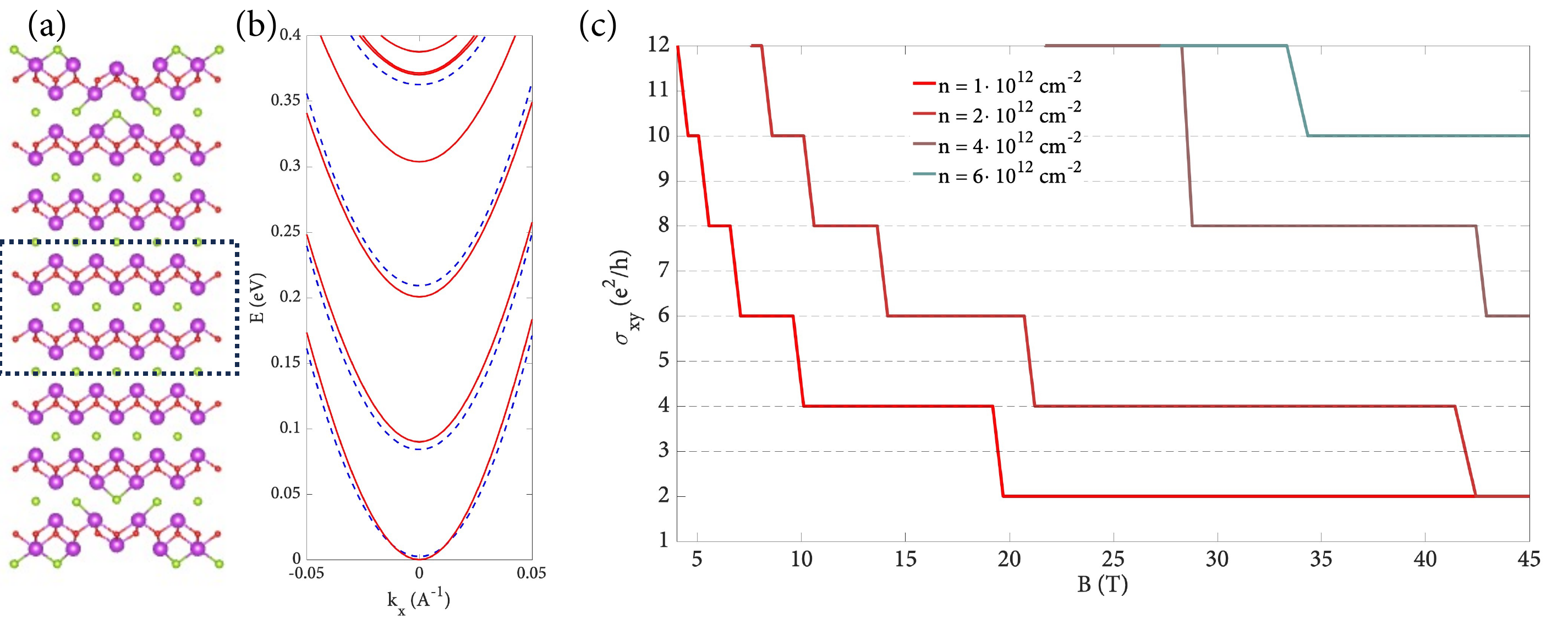}
    \caption{Quantum hall effect in few-layer Bi\textsubscript{2}O\textsubscript{2}Se with surface vacancies. (a) Band structure of a relaxed $4 \times 1$ 4 unit cell with $50 \%$ Se loss on the surface. Dashed lines denotes the unit cell that matches the description given in Sec.~\ref{sec:multilayer}. (b) Ab-initio band structure and effetive model fitting. Blue (dashed) line corresponds to the ab-initio data, red is the effective model. Fitting parameters are presented in the main text. (c) Quantum Hall effect in the 4 unit-cell system with ab-initio fitted parameters. For all densities, we find \textit{even integer} plateaus of $e^2/h$ while maintaining a single peak in quantum oscillations, indicating a single Fermi surface. Here $\Gamma = 11K$.}
    \label{fig:fig13}
\end{figure}
We carried out first-principle calculations using VASP \cite{VASP} on slabs of Bi\textsubscript{2}O\textsubscript{2}Se, while taking into account the experimental finding that nearly $\sim 50\%$ of the Se on the Se-terminated surface is lost to vacancies. Each reconstructed unit cell retains inversion symmetry. In Fig.~\ref{fig:fig13}(a) we show the structure of a reconstructed, relax $4\times1$ unit cell of Bi\textsubscript{2}O\textsubscript{2}Se with $50\%$ of Se content removed to account for vacancies. In Fig.~\ref{fig:fig13}(b) we fit a $4$ unit cell Hamiltonian effective model from Eq.~\eqref{eq:Hmulti} to ab-initio results (dashed blue line) to the parameters of the model which include $t,t', m^{*}, \alpha$. We find the best fit for $t=t' = 0.2\mathrm{eV}$ and $m^{*} = 0.14m_0$, $\alpha = 1.45 \mathrm{eV \AA}$. This value is in agreement with the relative magnitude of atomic SOC in Bi \cite{Kurita2020}, and is further reinforced by the fact that the character of the conduction states is exclusively Bi-$p_z$. 
\end{document}